\documentclass{jfm}

\usepackage{graphicx} 
\usepackage[utf8]{inputenc}
\usepackage{color} 
\usepackage{ulem}
\usepackage{amsmath} 
\usepackage{gensymb}
\usepackage{natbib}
\usepackage[lofdepth,lotdepth]{subfig}

\newcommand{\fig}[1]{Fig.~\ref{#1}}

\title[Internal Wave Attractors in 3D Geometries:  a dynamical systems approach]
{Internal Wave Attractors in 3D Geometries : a dynamical systems approach.}

\author[Grimaud Pillet, Leo Maas and  Thierry Dauxois]
  {Grimaud Pillet\textsuperscript{1}, 
  Leo R. M. Maas\textsuperscript{2}     and  Thierry Dauxois\textsuperscript{1}
   }
\affiliation{
  \textsuperscript{1}
 Université de Lyon,  ENS de Lyon, UCBL, CNRS, Laboratoire de Physique, Lyon, France\\
  [\affilskip]
 \textsuperscript{2}
  Institute for Marine and Atmospheric Research Utrecht, Utrecht University, Netherlands\\
}

\pubyear{???} \volume{???} \pagerange{???}
\date{\today}
\begin{document}

\maketitle

\begin{abstract}
We study the propagation in three dimensions of internal waves using ray tracing methods
and traditional dynamical systems theory.
The wave propagation on a cone that generalizes the Saint Andrew's cross justifies the introduction of an
angle of propagation that allows to describe the position of the wave ray in the horizontal plane.
Considering the evolution of this reflection angle for waves that repeatedly reflect off an inclined slope, 
a new trapping mechanism emerges that displays the tendency to align this angle with the upslope gradient.

In the rather simple geometry of a translationally invariant canal, we show first that
this configuration leads to trapezium-shaped attractors, very similar to what has been extensively studied in two-dimensions.
However, we also establish a direct link between the trapping and the existence of two-dimensional attractors.

In a second stage, considering a geometry that is
not translationally invariant, closer to realistic configurations, we prove that although there are no two-dimensional attractors, one can find
a structure in three-dimensional  space with properties similar to internal wave attractors: a one-dimensional attracting manifold. Moreover, as this structure is unique, 
it should be easy to visualize  in laboratory experiments since  energy injected in the domain would eventually be 
confined to a very thin region in three-dimensional space, for which reason it is called a super-attractor.
\end{abstract}


\section{Introduction}

The unusual properties of internal waves propagating in stratified fluids lead to a particularly interesting phenomenon that
has been studied in several situations in two dimensions: the existence of internal wave attractors.
The latter  correspond to a limit cycle towards which internal waves will focus in most confined geometries with at least one sloping boundary.
These beautiful mathematical patterns exist thanks to the very peculiar non-specular reflection law that linear  internal gravity waves obey.
Nonlinearity is effectively introduced by the dynamical system that repeated application of this reflection law entails.

The large majority of internal wave attractor studies were restricted to two-dimensio\-nal geometries~\citep{Maas2005}.
Despite their intriguing properties at the origin of many very interesting works,
if one wants to study their possible oceanographic and astrophysical relevance, one has to consider three-dimensional situations.
This is the main goal of this paper, in which we will use ray tracing to study the most interesting situations.

In the literature, wave attractor studies set in three dimensions
were mostly restricted to spherical shells~\citep{Rabitti2013}  that are particularly relevant in astrophysics \citep{RGV2001}.
Indeed, it is usual to consider the interior of gaseous planets as a fluid, (at least partially) stratified radially, around a solid or very dense liquid core~\citep{Dintrans}. 
Internal waves propagating within this spherical shell may indeed strongly influence the dynamics of the planet~\citep{Andre}. 
With a view of modelling internal tides in a channel, their three-dimensional behaviour was also investigated numerically in a rotating, uniformly-stratified parabolic channel  ~\citep{DM2007}. Taking into account the strong analogy between inertial and internal waves, one can also refer to
works that considered rotating fluids in spherical~\citep{Rabitti2014}, or trapezoidal ~\citep{MandersMaas2004} basins.

Using three dimensional ray tracing algorithms in these geometries, these authors were able to show that 
wave attractors obtained in two dimensions were not affected by the third dimension. They just
keep their two-dimensionality. Owing to a residual symmetry, such as an invariance to translation (due to along-slope uniformity) or rotation (cylindrical or spherical symmetry), a \textit{set} of attractors may exist side-by-side.  Together they can be seen as a two-dimensional attracting manifold.

One may however ask two important questions:
\begin{itemize}
\item[i)] What are the conditions for the existence of two-dimensional attracting manifolds when considering general three-dimensional geometries?
\item[ii)] Is it possible for basin geometries in which residual symmetries are absent to exhibit one-dimensional attracting manifolds, not contained in a plane? 
\end{itemize}

These are the two objectives of this paper.
In section 2,  we  first  present the propagation of internal wave beams before focusing on the reflection in  three
dimensions. We derive the reflection law for internal waves reflecting off an inclined slope and, using dynamical systems theory, we discuss in detail the map linking a wave beam's incident angle relative to the direction of the bottom gradient to its reflected angle.    
In section 3,  we exhibit two-dimensional attractors in three dimensions 
before considering, for the first time in section 4, a fully tridimensional geometry with 'super-attractors'.
Finally,
in  section 5,  we  conclude  and  draw  some  perspectives.

\section{Propagation and reflection in three dimensions}

\subsection{Three-dimensional propagation}

In an inviscid and incompressible fluid, linearly stratified along the vertical  $z$-axis, internal waves correspond to perturbations
of the velocity, the pressure and the  density fields
\begin{eqnarray}
 \overrightarrow{V} = \overrightarrow{V_0} + 
\overrightarrow{v}, \quad P = P_0 + p ,\quad \varrho = \bar{\rho}(z) + \rho'(x, y, z, t)
\end{eqnarray}
in which $\bar{\rho}(z)$ is the unperturbed linear stratification and with
 $\vert v \vert
\ll \vert V \vert$, $|p| \ll P_0$ and $|\rho'| \ll \rho_0$. 

\begin{figure}
\centering\null\hfill
\includegraphics[scale=1]{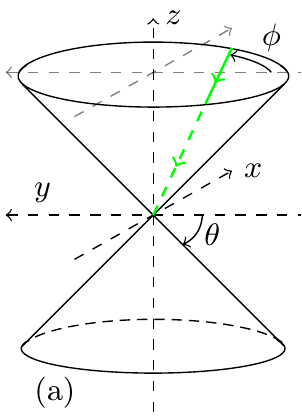}
\hfill
\includegraphics[scale=1]{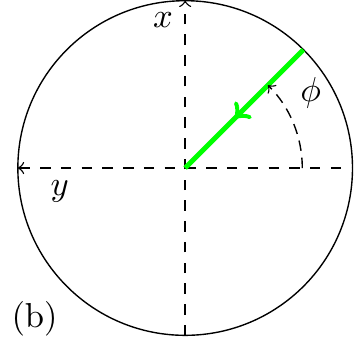}
\hfill\null
\caption{Double cone with aperture $\pi/2-\theta$ representing all possible propagating rays with frequency~$ \omega  = \pm N\sin \theta$.  
Either half of the double cone on one side of the apex is called a cone.
Drawn in green, a  ray is characterized by its horizontal angle $\phi$. Perspective view (a) 
and top view (b).}
\label{cone_prop}
\end{figure}

In the framework of the Boussinesq approximation with 
$\rho_0 = \langle \bar{\rho}\rangle$ the average density over the stratified region,
the projections of the Navier-Stokes equation on the three axes lead in the linear regime to 
\begin{eqnarray}
\dfrac{\partial \overrightarrow{v}}{\partial t} & = &  - \dfrac{1}{\rho_0}\overrightarrow{\nabla} p  +b\, \overrightarrow{e_z} \label{eq_vectoriel}
\end{eqnarray}
where buoyancy $b\equiv - \rho' g/\rho_0 $, while the conservation of buoyancy reads
\begin{eqnarray}
\dfrac{\partial b}{\partial t}+w N^2 & = & 0. \label{denseq}
\end{eqnarray}
Here we introduced the square of the buoyancy frequency 
$N^2 = -({g }/{\rho_0 }){\mbox{d} \bar{\rho}}/{\mbox{d} z}$, assumed to be constant. 
 Combining appropriately Eqs.~(\ref{eq_vectoriel}) and~(\ref{denseq}) and their time or spatial derivatives,
 one finally gets using the incompressibility condition
\begin{equation}
\dfrac{\partial^2}{\partial t^2} \nabla^2 v_z   + N^2\left(\dfrac{\partial^2}{\partial x^2} + \dfrac{\partial^2}{\partial y^2}\right)v_z = 0.
\label{eq_f}
\end{equation}
Plane waves  $v_z = v_{z0}\,
\exp(i(\omega t -\vec{k}.\vec{x}))$
 with wavevector $\vec{k}= (k_x,k_y, k_z)$ and frequency~$\omega$, are solutions provided that the dispersion relation 
\begin{eqnarray}
{\omega}{} & = &\pm N\sqrt{\dfrac{k_x^2 + k_y^2 }{k_x^2 + k_y^2 + k_z^2} }
\end{eqnarray}
is satisfied.
Plane internal waves of frequency $\omega$ propagate their energy along the direction of the group velocity vector. Let us denote  such an internal wave beam  a 'ray'. 
Calling $\theta$ the angle of the  ray with respect to the horizontal, 
one thus recovers in this three-dimensional setting $ \omega  = \pm N\sin \theta$ 
that requires that plane waves with frequency~$\omega$  are propagating with a given and constant angle $\theta$ with  the horizontal. 
In three dimensions, internal wave rays  of fixed frequency therefore lie on a double cone. 
As the current velocity vector associated with such an internal wave is parallel to the rays, these also lie on the same double cone
\begin{equation}
v_z = \pm \, \tan \theta \, \sqrt{v_x^2+v_y^2} \label{equationconeavectheta}
\end{equation}
represented in Fig.~\ref{cone_prop}(a) and that reduces to the St Andrew's cross in two dimensions.
The propagation of an internal wave ray with frequency $\omega$ is thus described by its 
position ($x$,$y$,$z$) and two angles: the horizontal propagation 
angle~$\phi$ with respect to the downslope-directed $y$-axis and the angle $\theta$ that 
is linked to its vertical inclination.

It is important to realise that these five parameters $(v_x,v_y, v_z,\theta,\phi)$ are not strictly equivalent to the
three components of the position and of the velocity given usually for describing the motion of an object.
 However, as we will
show below, due to focusing/defocusing of the wave component propagating in bottom-normal direction this norm is not conserved upon reflection of an internal wave off an inclined slope. During focusing it amplifies which increases our interest in its location. When a ray approaches an attractor, the ray path tells us where we should be looking because all the internal wave energy goes to that location.
Therefore, for now the main interest of this paper is in the direction of the ray, not the wave field's magnitude.

For any $\theta$-value, by stretching or compressing the $z$-axis with a factor $\tan \theta$, it is possible to map
the problem to the value $\theta = 45\degree$ that we will assume for the remainder of the paper. 
In this case, the three components of the velocity field of an internal wave ray characterized by the horizontal angle 
 $\phi$ are ($v_x$, $v_y$, $v_z $) = ($v_z \sin \phi$, $-v_z \cos 
\phi$, $v_z $), that leads precisely to the equality $v_z^2 = v_x^2 + v_y^2$.

\subsection{Reflection off an inclined plane}\label{Reflectiononaninclinedplane}

The impermeability condition when internal waves reflect from
a plane, $z=sy$ inclined with an 
angle $\alpha$ with respect to the horizontal $y-$direction and having slope $s=\tan \alpha$
 implies that the normal group velocity should vanish at the boundary,
while the inviscid hypothesis and the invariance along the $x$-direction leads to the conservation of the along slope group velocity.
Taking into account the incident (denoted with the index $i$) and reflected (denoted with the index~$r$) particle velocity  $\vec{v}_i \exp[ i( \omega_i t - \vec{k}_i\cdot\vec{r}) ] + \vec{v}_r \exp[ i( \omega_r t - \vec{k}_r\cdot\vec{r}) ]$,
the above condition first implies that the frequency~$\omega$ is kept constant.
Introducing the incident  $\vec{v}_i =(v_{x,i}$, $v_{y,i}$, $v_{z,i})$ and the reflected 
$\vec{v}_r = (v_{x,r}, v_{y,r}, v_{z,r})$  
velocity fields, these remarks can be summarized in the following three conditions for reflection:
\begin{itemize}
\item[$(a)$] $\theta_r =\pm \theta_i \,  \mbox{mod}\, \pi$: the reflected and incident waves are on the same double cone.
\item[$(b)$] $v_{x,r}$ = $v_{x,i}$: the along-slope component is unchanged.
\item[$(c)$]  The component of the velocity normal to the slope changes its sign, that leads to \begin{eqnarray}
v_{z,r}\cos \alpha      - v_{y,r}\sin \alpha     & = & -(v_{z,i}\cos 
\alpha    -  v_{y,i}\sin \alpha ) \end{eqnarray}
that can be simplified using $s = \tan \alpha $ as \begin{eqnarray}
v_{z,r}  + v_{z,i} & = & s(v_{y,r}  +  v_{y,i}). \label{equationavecsaciter}
\end{eqnarray}
\end{itemize}
Condition $(a)$ taking into account Eq.~(\ref{equationconeavectheta}) with $\theta=45\degree$
leads directly to 
$v_{z,r}^{2}  =  v_{x,r}^{2} + v_{y,r}^{2}$ and $v_{z,i}^{2} = v_{x,i}^{2} + v_{y,i}^{2}$.
Subtracting both equalities and using condition $(b)$, one gets
\begin{eqnarray}
(v_{z,r}-v_{z,i})(v_{z,r}+v_{z,i}) & = & (v_{y,r}-v_{y,i})(v_{y,r}+v_{y,i}) .\label{equationavecsaciterbis}
\end{eqnarray}
For a sloping boundary,  $(v_{y,r}  +  v_{y,i})$ is non-zero. Hence, by combining Eqs.~(\ref{equationavecsaciter}) and~(\ref{equationavecsaciterbis}) this term can be divided out, which
leads to the following system of two equations with two unknowns  $v_{z,r}$ and $v_{y,r}$:
\begin{eqnarray}
v_{z,r} + v_{z,i}  =  s (v_{y,r} + v_{y,i}),\label{neweqa}\\
v_{y,r} - v_{y,i}  =  s (v_{z,r} - v_{z,i}). \label{neweqb}
\end{eqnarray}
Recalling condition $(b)$ for the along-slope velocity component, the above system leads to 
\begin{eqnarray}
v_{x,r}  & =  & v_{x,i},\label{equationpourvx}\\
v_{y,r}  & =  & \dfrac{(1+s^2)v_{y,i} - 2sv_{z,i}}{1-s^2} \label{loi_vit}, \\
v_{z,r} & = & \dfrac{-(1+s^2)v_{z,i} + 2sv_{y,i}}{1-s^2}, \label{equationpourvz}
\end{eqnarray}
providing the reflected ray from the knowledge of the incident one.

It is interesting  to derive the corresponding law between the incident and reflected horizontal angles $\phi_i$ and 
$\phi_r$, both angles being defined in \fig{th_ref}. 
Recalling that $v_x=v_z \sin \phi$ and $v_y=-v_z \cos 
\phi$, the ratio of Eqs.~(\ref{equationpourvx}) and (\ref{equationpourvz}) 
\begin{eqnarray}
\dfrac{v_{x,r}}{v_{z,r}}=\sin \phi_{r}   & = &  \dfrac{(s^2-1)\sin \phi_i}{1+s^2 
+ 2s\cos 
\phi_i },
\label{loi_sin}
\end{eqnarray} 
$\phi_i'={\cal R}(\phi_i,s)$, detailed in appendix A.
Above formula was originally derived by~\cite{Maas2005}. Interestingly, one can easily check that this law is also valid
when considering horizontal {($s=0$)} or vertical {($s{\rightarrow}\infty$) boundaries.

Note that in the remainder of the paper, this transformation will always be associated 
with an appropriate normalisation of the 
reflected velocity field.
Indeed, as in two dimensions, the norm of the velocity field is not a conserved quantity through the reflection mechanism. Consequently, such 
a normalisation is natural to avoid any
numerical difficulties, especially as our primary interest is the direction of the velocity vector rather than its magnitude.

\subsection{Study of the reflection map ${\cal R}$}

In order to understand the variation of the angle because of the reflection,
instead of studying the reflected angle $\phi_r$ itself, it is more appropriate to 
consider $\phi_i^{'}$, the following incident angle once the ray will come back towards the slope upon an intermediate surface reflection,
analogously to a Poincar\'e map.
Indeed, after reflection the green ray represented in \fig{th_ref} is transformed into the red one that will encounter horizontal or
vertical walls, before coming back towards the slope of interest, with an angle $\phi_i^{'} = \pi +  
\phi_r$ as shown by the green dashed ray in the example of \fig{th_ref}b.
An example of such a possible trajectory is depicted in~\fig{sketch_attract}(a). 

This is of course not the general situation since the ray can impinge on a vertical wall and then on an horizontal one,
before coming back towards the slope with an angle $\phi_i^{'} = -\phi_r$. However, let us discuss first a simple case
that will allow us to derive global and rather generic properties of this reflection law. 

 \begin{figure} 
\centering
\null\hfill
\subfloat{
\includegraphics[scale=1]{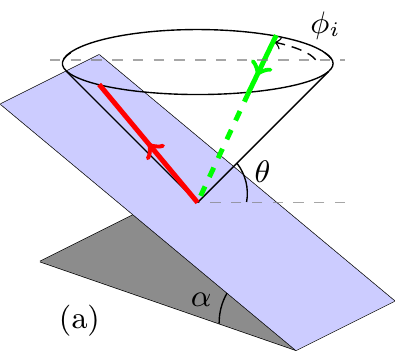}
}
\hfill 
\subfloat{
\includegraphics[scale=1]{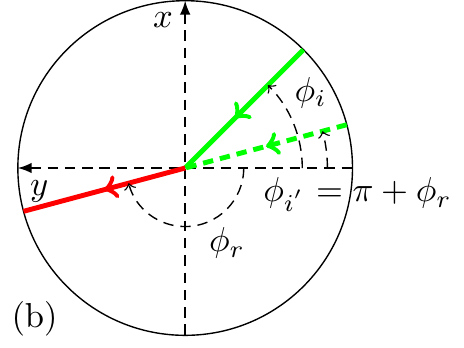}
}
\hfill\null
\caption{Perspective view (a) and top view (b) of the reflection of an internal wave beam off an inclined slope. 
The bottom, inclined at angle $\alpha$ with respect to the horizontal $xy$-plane, is represented by the inclined 
blue rectangle. The internal wave beam propagates along a cone whose inclination, $\theta$, is set by the 
ratio of wave and buoyancy frequencies. The incident (in green) and reflected (in red) beams make angles 
$\phi_i>0$ and $\phi_r<0$ relative to the downslope direction, respectively. The green dashed line is discussed in the text. }
\label{th_ref}
\end{figure}

 \begin{figure}
\centering
\includegraphics[scale=0.93]{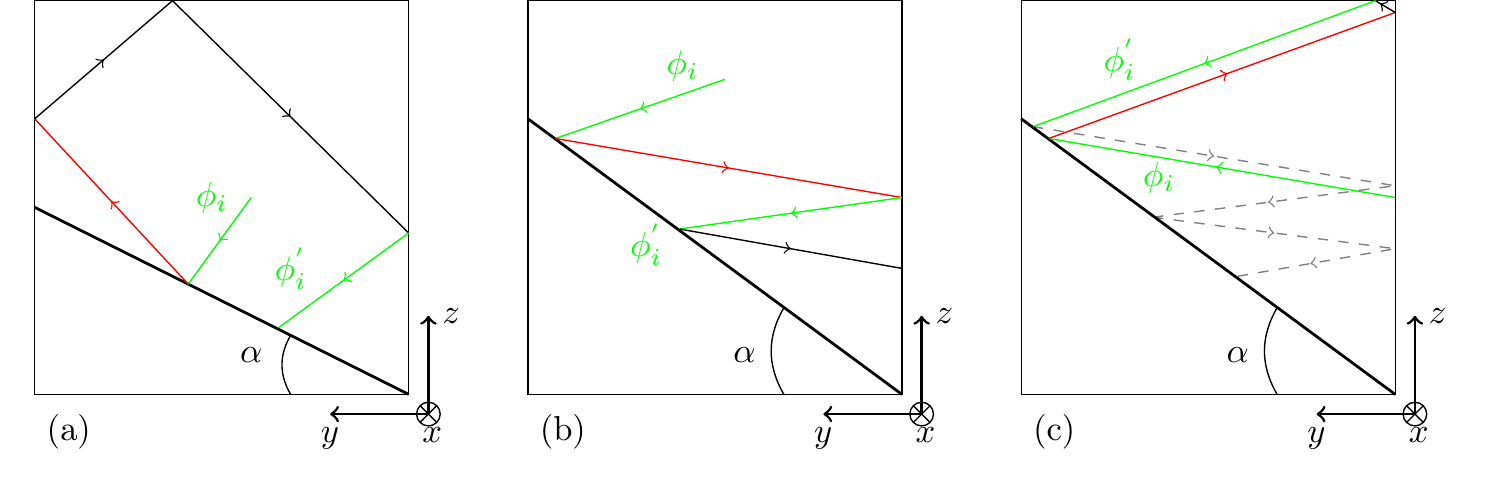}
\caption{Possible reflections of an internal wave off a bottom whose slope is  subcritical (left panel) or
supercritical (center and right panels).  In the right panel, the trajectory is represented with a dotted line after the second reflection. 
Next to each ray, the corresponding horizontal angle $\phi$ is written. Note that this picture, being a side view displays only the projection  of horizontal angles $\phi$ on the $yz$ plane. }
\label{sketch_attract}
\end{figure}

Once the angle of the slope is given, the value of  $s$ is known and then one can study the map~${{\cal R}}$ providing $\phi_i^{'}$ 
as a function of $\phi_i$ (see Appendix A for useful detail). Note that for $s=\tan \alpha<1$, only the top cone  of~Fig.~\ref{cone_prop} is physically interesting since 
the bottom one is fully below the slope. 

Let us study separately the three different cases: $s <1$, $s=1$ and  $s>1$.

\subsubsection{{Sub}critical reflections $s<1$}

Figure~\ref{ptfixe_sur} presents the evolution of the angle~$\phi_i^{'}$ as a function of $\phi_i$, for different subcritical cases with $s<1$. 
One realizes immediately that for any $s$-value, the function intersects the diagonal line in $\phi_i=0$ and $\pm\pi$. There are therefore three fixed points of map~${{\cal R}}$ (actually two since $\pm\pi$ correspond to the same physical state).
For any value $s<1$, as ${{\cal R}}'(0) < 1$ while ${{\cal R}}'(\pm\pi) > 1$, only $\phi^\star=0$ is a stable fixed point, corresponding to upslope propagation. 
The reflection off the inclined slope of an internal wave ray has therefore the systematic tendency 
to reduce its horizontal angle. This is what we will call in the remainder of the paper,
the trapping effect. Figure~\ref{ptfixe_sur}  shows that the closer the value of  $s$ is to 1, the faster is the trapping effect.
\begin{figure}
\centering
\includegraphics[scale=0.8]{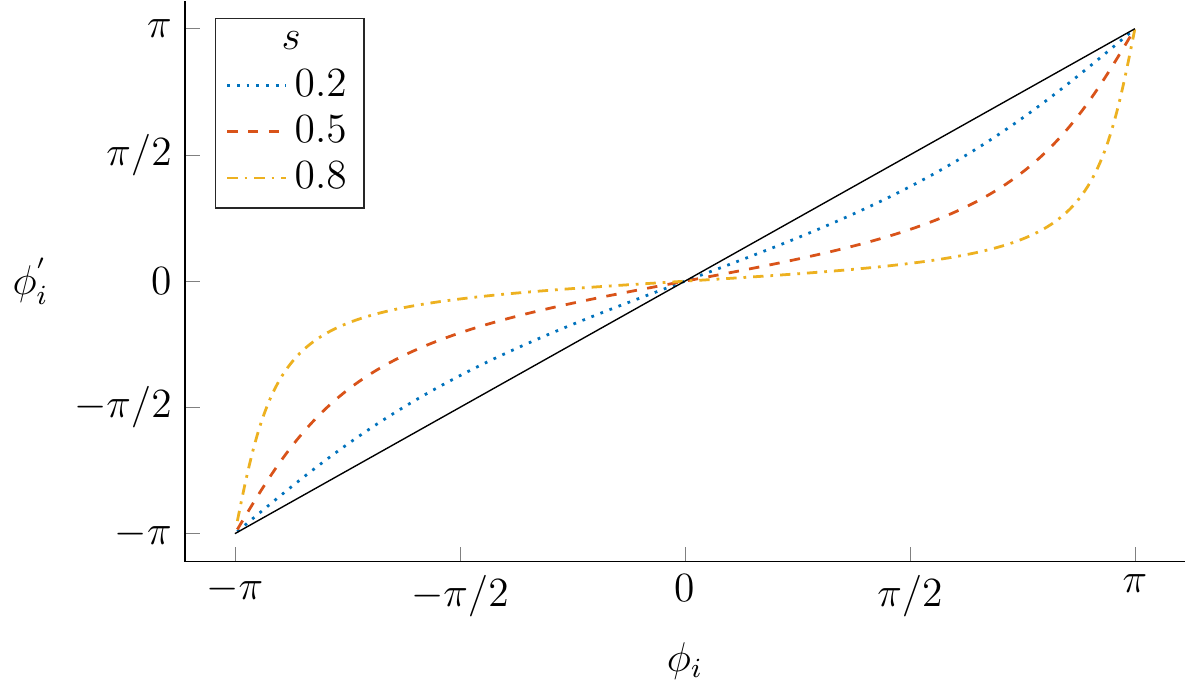}
\caption{Map $\phi_i' = {{\cal R}}(\phi_i,s)$ for $\phi_i \in [-\pi,\pi]$ for three different values of~$s<1$. 
The solid diagonal line represents  $\phi_i^{'}=\phi_i$.}
\label{ptfixe_sur}
\end{figure}

\subsubsection{Critical reflection $s=1$}

In the theoretical case $s=1$, the angle of the slope coincides with the aperture of the double cone.
Equations~(\ref{loi_vit}) and~(\ref{equationpourvz}) show that, in that case,  $v_{y,r}$ and $v_{z,r}$ diverge. Equation~(\ref{loi_sin}) is however
still valid and leads to $\phi_i^{'} = 0$ for any initial horizontal angle~$\phi_i$. The trapping is total from the very first reflection.

\subsubsection{{Super}critical reflection  $s>1$}

Figure~\ref{cone_sous} shows that in that case the top cone is not anymore fully attainable. 
Introducing the limiting angle $\phi_\ell = \arctan \sqrt{s^2-1}$, one realizes that $\phi_i $ has to be restricted to 
$[-\pi+\phi_\ell,\pi-\phi_\ell]$ for the top cone and to 
$[-\phi_\ell,\phi_\ell]$ for the bottom one.  Figure~\ref{ptfixe_sous} 
presents the analysis for reflections on the top ($v_{z,i} < 0$) and the bottom ($v_{z,i} >0$) cones, that have to be studied
separately.

\begin{figure} 
\centering
\null\hfill
\subfloat{
\includegraphics[scale=0.3]{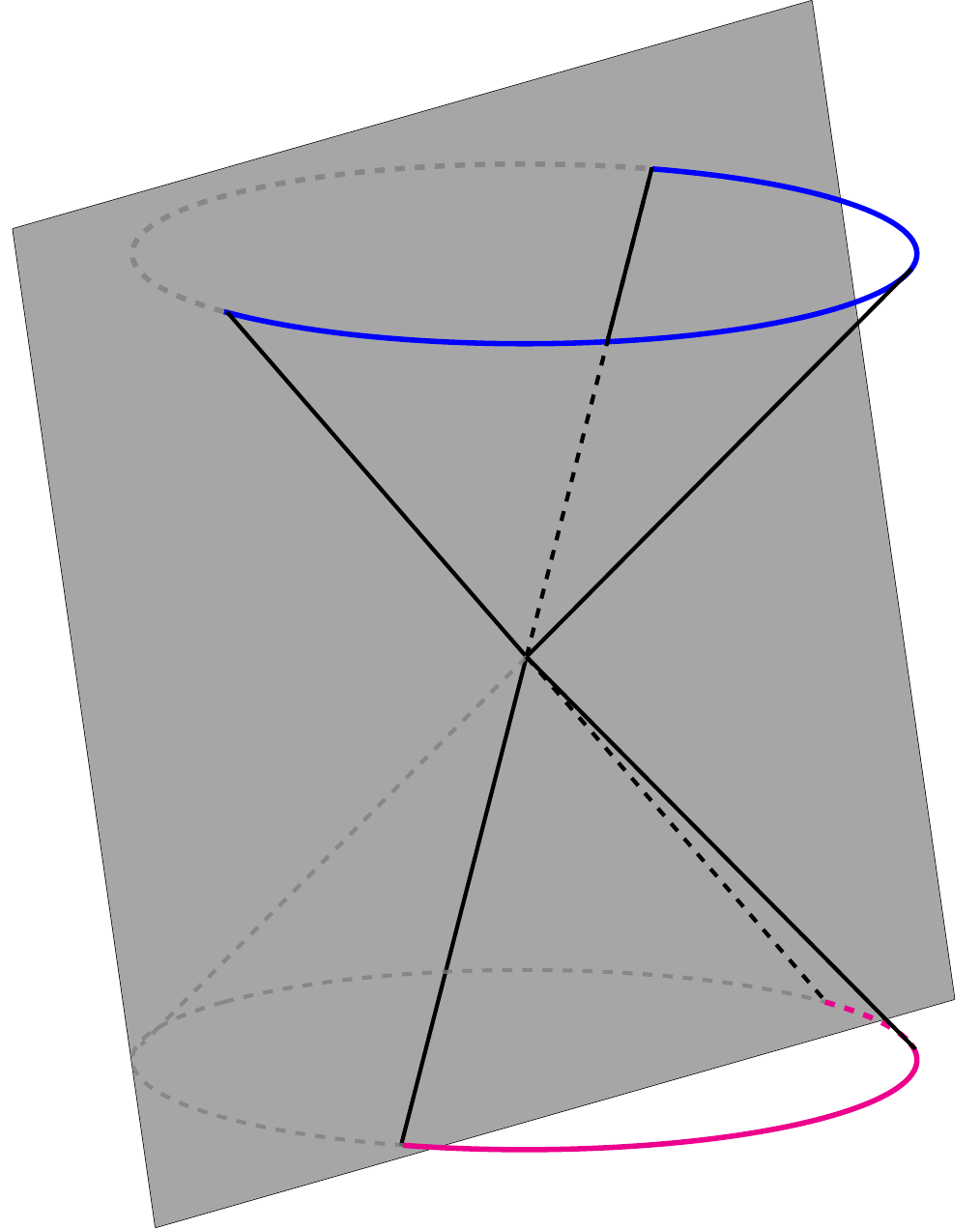}
}
\hfill 
\subfloat{
\includegraphics[scale=1]{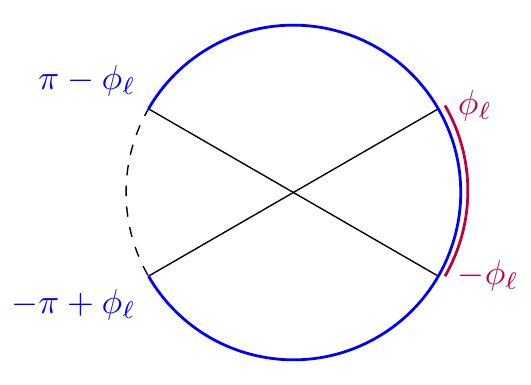}
}
\hfill\null
\caption{Cone of possibilities for a \textit{super}critically sloping bottom $s>1$. The left panel shows that the planar slope, in grey, inclined with
an angle $\alpha$, partially intersects the double cone. The right panel presents the available angular sector for 
$\phi_i$ on the top  (resp. bottom) cone in blue (resp. mauve). }
\label{cone_sous}
\end{figure}

\begin{itemize}

\item[i)] For rays impinging on the top cone, $\phi_i \in 
[-\pi+\phi_\ell,\pi-\phi_\ell]$, one gets  $v_{z,r}<0$, implying that the reflected ray belongs (and is restricted) to the bottom cone. 
 
Reflections on vertical or horizontal boundaries leading to $\phi_i^{'} = \pi + \phi_r$, 
like in Fig.~\ref{sketch_attract}(a), are therefore not possible. On the contrary, the situation is like the one
presented in~\fig{sketch_attract}(b), in which the wave reflects on a vertical boundary
before coming back towards the slope, leading to $\phi_i' = -\phi_r$ shown in
  \fig{ptfixe_sous}. 
Here again, there is one unique stable fixed point, that is $\phi^\star=0$. One finds once more
that the three-dimensional reflection law has the tendency to straighten the horizontal angle
of propagation.
 However, this case is of limited consequences; indeed, 
in the canal geometry that we will be interested in, {when $s>1$},
the two-dimensional version leads to a point
attractor, with all wave rays ending in the bottom right corner  as one can guess by looking at \fig{sketch_attract}(c).

\begin{figure}
\begin{center}
\includegraphics[width=1\textwidth]{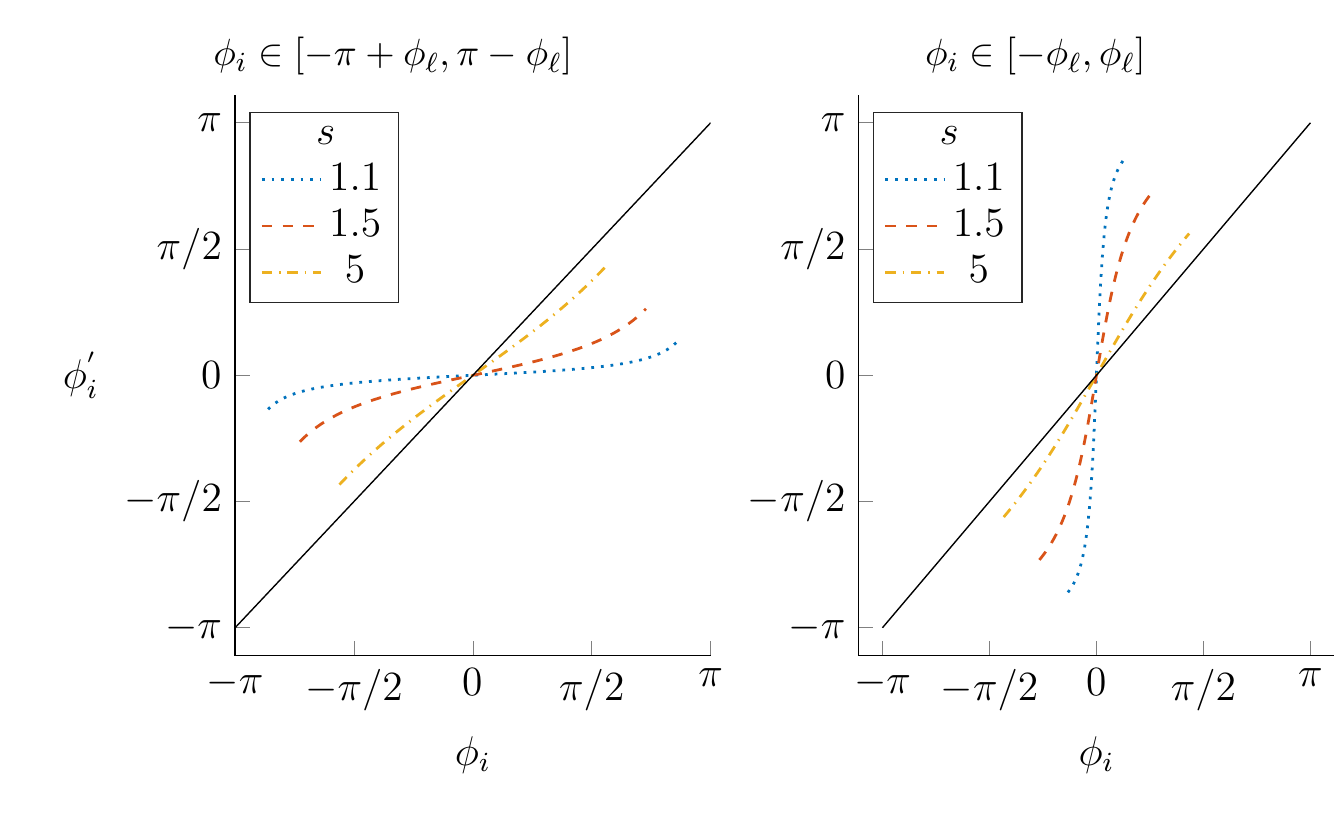}
\caption{
Map $\phi_i'= {{\cal R}}(\phi_i,s)$  
 for three different values of~$s>1$ for an internal
wave ray impinging on the top cone (left panel with $v_{z,i} = -1$) and  on the bottom one (right panel with $v_{z,i} = 1$). 
The solid diagonal line represents  $\phi_i^{'}=\phi_i$.}
\label{ptfixe_sous}
\end{center}
\end{figure}

 \item[ii)] For incident wave rays on the bottom cone, $\phi_i \in 
[-\phi_\ell,\phi_\ell]$ that leads to $v_{z,r} > 0$: the reflected ray is therefore on the top cone. 
Not unexpectedly, this case corresponds precisely to the previous one, 
once the sense of propagation has been reversed.
After a series of reflections as plotted in~\fig{sketch_attract}(c), one 
 finds the reflection law ${{\cal R}}$ detailed in Appendix A for $v_{z,i}>0$ (see right panel of \fig{ptfixe_sous}). 
This time, the reflection is defocusing since $|\phi_i^{'}| > |\phi_i|$ 
for any value $s>1$: the fixed point $0$ is thus unstable. However, as shown by the example 
depicted in~\fig{sketch_attract}(c), one will not indefinitely get reflections with  $v_{z,i} > 0$
since, eventually, due to a reflection at the horizontal rigid-lid surface one will get  $v_{z,i} 
< 0$, corresponding to the previous case. 

Moreover, if the upper boundary of the domain (the surface in the present case) is high enough,
as the function is an increasing function of its argument, 
one eventually reaches a value $\phi_i'$ 
that will be greater than $\phi_\ell$. Above that value, one cannot have anymore
reflection with $v_z > 0$, and one comes back to the previous case converging towards 
the fixed point~$\phi^\star = 0$.

\end{itemize}

In summary, in the situations that we have considered, the ray is eventually trapped in the plane corresponding to a vanishing horizontal
 angle $\phi$. This is the generic case, but as we will discuss below, trapping may not occur in some peculiar cases.
 
\subsubsection{Focusing or trapping?} 
 
We called {\it trapping} the alignment of the horizontal angle $\phi$ with respect to the downslope  direction of the reflecting slope. 
It is important to distinguish this from the focusing that occurs in two-dimensions, when a corresponding  
 internal wave beam reflects off a slope. Focusing corresponds to the decrease of the width of the beam after reflection,
or alternatively, to the decrease of the distance between two rays initially parallel and impinging on a planar slope.

\begin{figure}
\begin{center}
\includegraphics[width=0.9\textwidth]{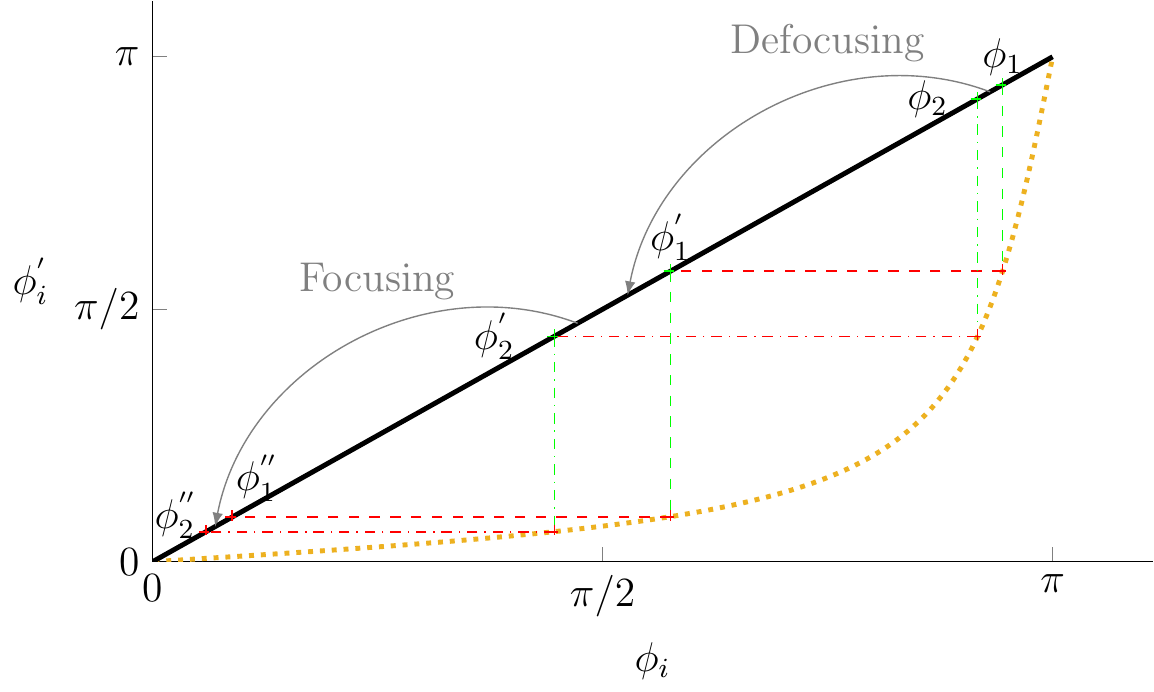}
\caption{
Map $\phi_i'={{\cal R}}(\phi_i,s)$ for $\phi_i \in [0,\pi]$ 
for~$s=0.8$ represented with the yellow dotted line. We follow two different reflections (dash-dotted and dashed lines) initiated from 
two initial horizontal angles $\phi_1$ and 
$\phi_2$. As previously, green (resp. red) corresponds to incident (resp. reflected) rays.
 The  diagonal line $\phi_i^{'}=\phi_i$
is represented by the solid black line.}
\label{foc_pie}
\end{center}
\end{figure}

In three dimensions, the analog of this focusing would correspond to a study of the angular gap
 $\phi_2 - \phi_1$ between both rays. In a situation where trapping is present as shown in~\fig{foc_pie} for $s=0.8$, one
 sees that the reflection off an inclined slope
could be focusing or defocusing.
The gap between two initial angles
increases after the first reflection, before decreasing. The first reflection
is therefore defocusing, while the second is focusing.
On the contrary, both reflections lead to smaller values of the angle,
they are trapping. Trapping and focusing are corresponding therefore to different ideas.
Defocusing occurs between two rays characterized by $\phi_1$ and $\phi_2$ if and only if
${{\cal R}}'(\phi)>1$ for  $\phi\in[ \phi_1$,$\phi_2$],where a prime indicates a derivative to its argument. Reciprocally,  focusing occurs
when ${{\cal R}}'(\phi)<1$ in this interval.

\section{2D attractors in a 3D geometry}
\label{canal}

\subsection{Choice of the geometry}

\begin{figure} 
\centering
\includegraphics[scale=1]{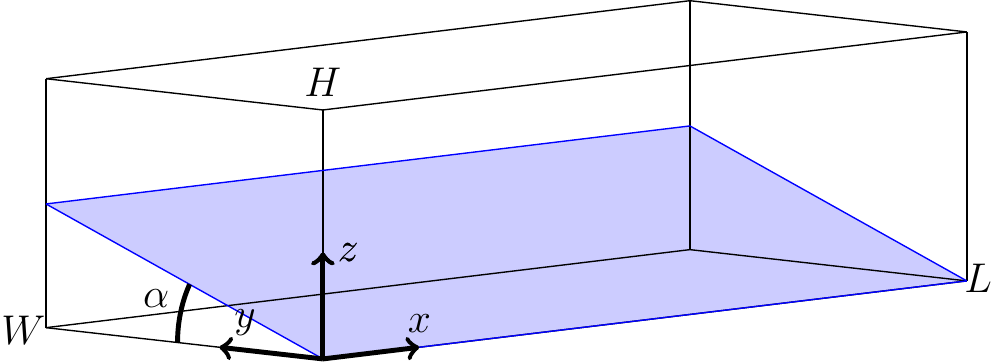}
\caption{Geometry under investigation with the definition of the height $H$, length~$L$ and width $W$ of the canal.
The slope, inclined at the angle $\alpha$ with respect to the horizontal $xy$-plane, is represented by the blue rectangle.}
\label{geom}
\end{figure}

Having presented the trapping mechanism due to the reflection off a slope, let us turn to its consequences
in a canal with an inclined slope as depicted schematically in~\fig{geom}. 
This geometry, that we used to derive the reflection law~(\ref{loi_vit}), has several advantages:
\begin{itemize}
\item[i)]  
Corresponding to a schematic simplification of estuaries or river arms, it has {\it a geophysical interest}.  
The Lower St. Lawrence Estuary (Eastern Canada) with its river bed essentially U-shaped transversally and longitudinally invariant over 1000 km~\citep{ElSabh} is a prototypic example.  
This site is remarkable since even though internal tides
are known to be generated at the land-locked head of the Channel 
\citep{Cyr2015}, 
surprisingly low intensity internal tides have been measured near the mouth of the
Laurentian Channel, eastern Canada~\citep{SaintLuarentAttenuation}.
It is therefore important to study propagation and reflection of internal waves in such a
geometry.

\item[ii)]
This geometry is {\it simple and easy to implement experimentally}, before studying 
in a second stage more complicated ones.

\item[iii)] {\it A theoretical interest} can also be anticipated from this geometry
from the study presented  in the preceding section. Indeed, 
when an internal wave ray reflects off a subcritical slope
 (as sketched in~\fig{sketch_attract}(a)), there is one single fixed point of the iterated map, $\phi^\star=0$,
that proves that the ray will eventually converge to a $yz$-plane, transverse to the canal. 
The internal wave will therefore be trapped. 

\end{itemize}

Interestingly, the transversal cut of the canal is precisely the appropriate geometry leading, in two dimensions, 
to internal wave attractors. 
In a given geometry, an internal wave attractor is a path towards
which all internal waves of a given frequency will converge: the existence of
such a limit cycle has been tested through 
ray tracing and experiments in various geometries \citep{MaasLam1995,Maas2005,Brouzetetal2016,BrouzetPhD,PilletPhD}, and has been confirmed in an exceptional case analytically \citep{Maas2009}.
Depending on a dimensionless lumped parameter containing the aspect ratio and the  ratio of wave to buoyancy frequencies, for the same geometrical domain, different attractors exist; they are labelled using two indices ($m,n$),
in which $m$ and $n$ describe the number of reflections on a vertical wall 
and on the slope respectively. 

\subsection{Simple attractors}

Let us consider first the simplest case for which the transverse geometry (i.e. in the $yz$-plane)
leads to (1,1) attractors. We will moreover consider the subcritical case $s<1$.
As the successive reflections will occur with an incident horizontal angle $\phi_i$ between
$0$ and $\pi$, the right panel of \fig{ptfixe_sur} shows that the angle will converge towards the fixed point
$\phi^\star=0$. This is indeed possible since the  (1,1) attractor loop allows  
$\phi_i' = \pi + \phi_r$, as shown by~\fig{sketch_attract}(a).

The important parameters for ray tracing are:
\begin{itemize}
\item[$\bullet$]  Geometrical ones ($H$, $W$, $\alpha$). 
Note that the dimensionless length of the canal is  $L=1000$.
\item[$\bullet$]   The angle of propagation of internal waves 
$\theta$ 
has been chosen to lead to a (1,1) attractor. It is not useless to recall that, in the ray tracing,   $\theta$ is always equal to $\pi/4$ by modifying the height $H$ with the factor $\tan \theta$ 
that stretches the vertical.

\item[$\bullet$]  The initial values of the ray ($x_0$, $y_0$, $z_0$, $\phi_0$) 
and $v_{z0}$ that determines the sheet of the double cone that is initially chosen.
\end{itemize}

\begin{figure}
\begin{center}
\includegraphics[width=\textwidth]{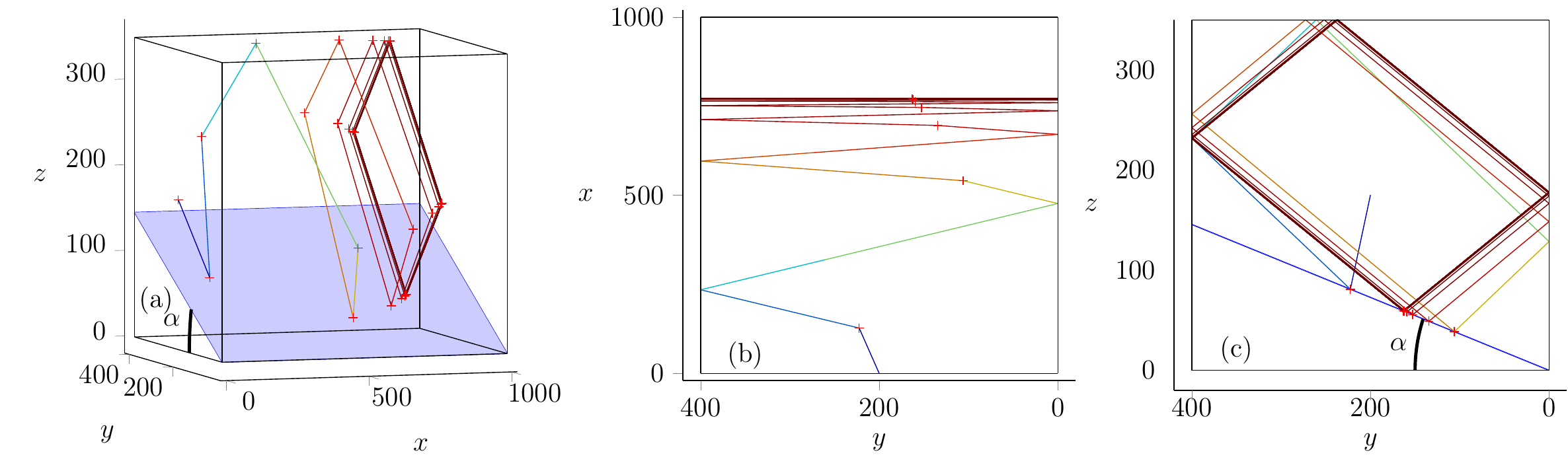}
\caption{
Perspective (a), top (b) and side (c) views of the trajectory of a single internal wave beam propagating in
the canal-like geometry filled with a linearly stratified fluid and the following
geometrical parameters: $H=350$, $W=400$, $L=1000$, $\alpha=20\degree$ and $\theta=35\degree$. 
The beam is sent downwards
(in the negative $z$-direction) from the plane $x=0$ with $y_0 = W/2$ and $\phi_0 \simeq \pi/2$ (i.e. into the positive along-tank $x$-direction).
Panel (b) shows the convergence of the horizontal angle
$\phi$ towards~$\phi^\star=0$, (rebounds on the slope are indicated with red crosses), while panel (c) emphasizes the limit cycle of the (1,1) attractor. 
The color of the ray progressively changes from blue to red with the advancement of the ray.
}
\label{foc_lente}
\end{center}
\end{figure}

A typical trajectory is plotted in  \fig{foc_lente} with different views.
It is clear that the ray, initially launched in the longitudinal $x$-direction,
after a finite number of reflections on the sloping bottom, 
eventually rotates towards a transverse plane.
Figure~\ref{foc_lente}(c) reveals that the transverse structure of the trajectory is an attractor,
identical to those obtained in 2D~\citep{MaasLam1995,Maas2005,Brouzetetal2016}. 
Note however a fundamental difference with respect to the 2D propagation: the rays do no longer
propagate only along one of the four different angles  $\theta$, $-\theta$, $\pi-\theta$, $\pi+\theta$, but involve a horizontal angle of propagation, $\phi$, too.   \fig{foc_lente}(c)
shows only the angle projected on the transverse plane; it coincides precisely with one of these four possibilities only if 
$\phi=0$, which indeed is reached asymptotically.

The above discussion has emphasized how the trapping mechanism occurs in three dimensions, and transforms an initial longitudinal propagation into an
attractor in the transverse plane. We will turn towards the trapping and the convergence times, two important notions.

\subsubsection{The trapping time}\label{trappingtime}

The speed of convergence of the trapping is not always as progressive as the example shown in
Fig.~\ref{foc_lente} for which, in order to identify the different regimes,
parameters have been tuned to get a trapping, neither too fast, nor too slow.

Instead of considering the evolution of the angle $\phi$ as a function of the number of reflections, 
one can study the longitudinal velocity component~$v_x$. 
As briefly discussed in Section~\ref{Reflectiononaninclinedplane}, because of the specifics of this reflection process, 
the longitudinal component $v_x$ stays constant while both
transverse ones, $v_y$ and $v_z$, diverge towards infinity. To avoid this divergence, the total velocity has therefore been  normalized
after each reflection. 
During the trapping, one thus gets $v_z \rightarrow \pm1/\sqrt{2}$ and $v_y \rightarrow \mp 1/\sqrt{2}$ while $v_x \rightarrow 0$,
the signs between $v_y$ and $v_z$ being exchanged at each reflection.
 
\begin{figure}
\begin{center}
\includegraphics[width=.7\textwidth]{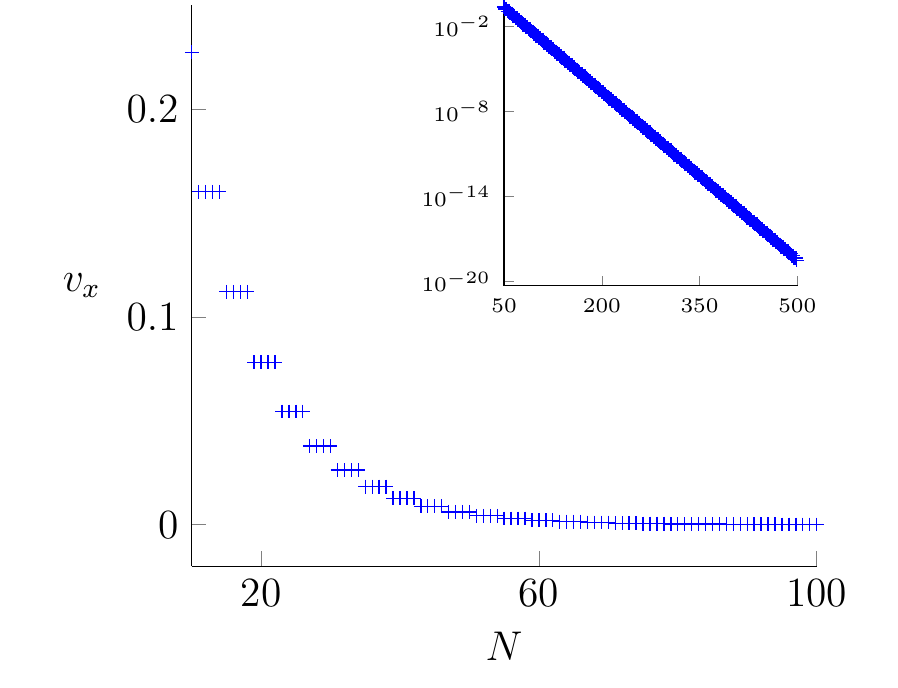}
\caption{Evolution of the longitudinal velocity component $v_x$ in linear scale as a function of the number of reflections.
The inset presents the same plot in semi-logarithmic scales. {For this plot $\alpha = 8 \degree$ and therefore $s= 0.18$.}}
\label{vx_lente}
\end{center}
\end{figure}
Figure~\ref{vx_lente} presents the longitudinal component $v_x$ as a function
of the number of reflections on boundaries.
The rebounds on the horizontal or vertical boundaries do not modify  $v_x$. On the contrary, 
it strongly decreases when the reflection occurs on the slope. 
In agreement with the geometrical structure of the (1,1) attractor with three non-focusing vertical or horizontal sides
and only one slope, this is the reason for the four identical values before a drop corresponding to a reflection off
the inclined slope.

To characterize the velocity of trapping that Fig.~\ref{vx_lente} suggests to be exponential, let us introduce 
the coefficient $\gamma_p$ defined as the ratio
between the horizontal velocity components before the reflection, $v_x$,  and after,~$v_{x}^{'}$.
Since the horizontal component is modified only by the normalization, one gets
\begin{eqnarray}
\gamma_p &=& 
\left(v_x^{'2}+v_y^{'2}+v_z^{'2}\right)^{-1/2}\\
 & = & \left(v_x^2 + \frac{1}{{(1-s^2)}^2}\left[ 
(v_y^2+v_z^2)((1+s^2)^2+4s^2) - 8s v_y v_z(1+s^2)  \right] \right)^{-1/2}.
\end{eqnarray}
In order to get rid of the dependence of this coefficient with respect to the components of the
velocity, it is necessary to consider the regime close to the convergence towards the fixed point.
As discussed in previous section, in this regime, the horizontal component~$v_x$ can be neglected with respect to 
$v_y$ and $v_z$. Taking advantage of the normalization, one gets $v_y^2+v_z^2 \approx 1$ and $v_yv_z = 
-1/2$, that leads to
\begin{eqnarray}
\gamma_p 
&\simeq&  \left(0 + \frac{\left[(1+s^2)^2+4s^2 +4s (1+s^2)  \right]}{{(1-s^2)}^2} \right)^{-1/2}\\
&=& \ \frac{1-s^2}{\left(1+4s+6s^2+4s^3 +s^4\right)^{1/2} }\\
&=& \dfrac{1-s}{1+s},
 \label{gamma}
\end{eqnarray}
the focusing power of normally-incident internal waves reflecting off an inclined wall of slope $s$.
In the framework of this approximation, the convergence is therefore exponential with the number~$N$ of reflections since one can write
$v_x(N) =  v_x(0) \, \gamma_p^N =   v_x(0) \, e^{N \ln \gamma_p}$. It is straightforward to check that $\gamma_p$ is less than 1;
it is also
a decreasing function with the gradient of
the slope,~$s$, and tends towards zero as  $s$ tends toward unity (we recall that this discussion is performed in the subcritical regime).

The example presented in \fig{vx_lente} attests that this approximation is very quickly valid. The inset of the exponential relaxation leads to
 $\ln \gamma_{p}^{num} \simeq -0.091$ that one can compare to the predicted value. Once the value  $s=\tan \alpha/\tan \theta=0.178$ is obtained
from the angle of the slope $\alpha=8\degree$, one just has to realize that only the reflection on the slope is effectively trapping,
while the three successive reflections on the vertical or horizontal boundaries are not modifying the velocity components.
Once this factor four is taken into account, one gets  $\ln \gamma_{p}^{th} \simeq -0.090$ that does confirm the approach.

\subsubsection{The convergence time} 

It is important to make a distinction between the trapping time, defined as the inverse of $\ln \gamma_p$, and
the convergence time, that could be defined as the time for the ray to be really trapped. Indeed, successive reflections
on vertical and horizontal boundaries do not necessarily each lead to $\phi_i^{'} = \pi +\phi_r<\phi_i$ as in the example  in \fig{sketch_attract}(a). 
Several untrapping reflections can follow one another, and consequently significantly delay the trapping.

Figure~\ref{ex_vx} presents three examples, in which only the initial position $y_0$ has been modified,
 but leading to significantly different convergence times. Geometrical parameters used in the ray tracing presented in~\fig{foc_lente} have
 been kept constant, but only the value for $\alpha$ is now smaller to get a slower trapping, and the value $\theta$ has been modified
 to correspond again to a (1,1) attractor.

\begin{figure}
\begin{center}
\includegraphics[width=1\textwidth]{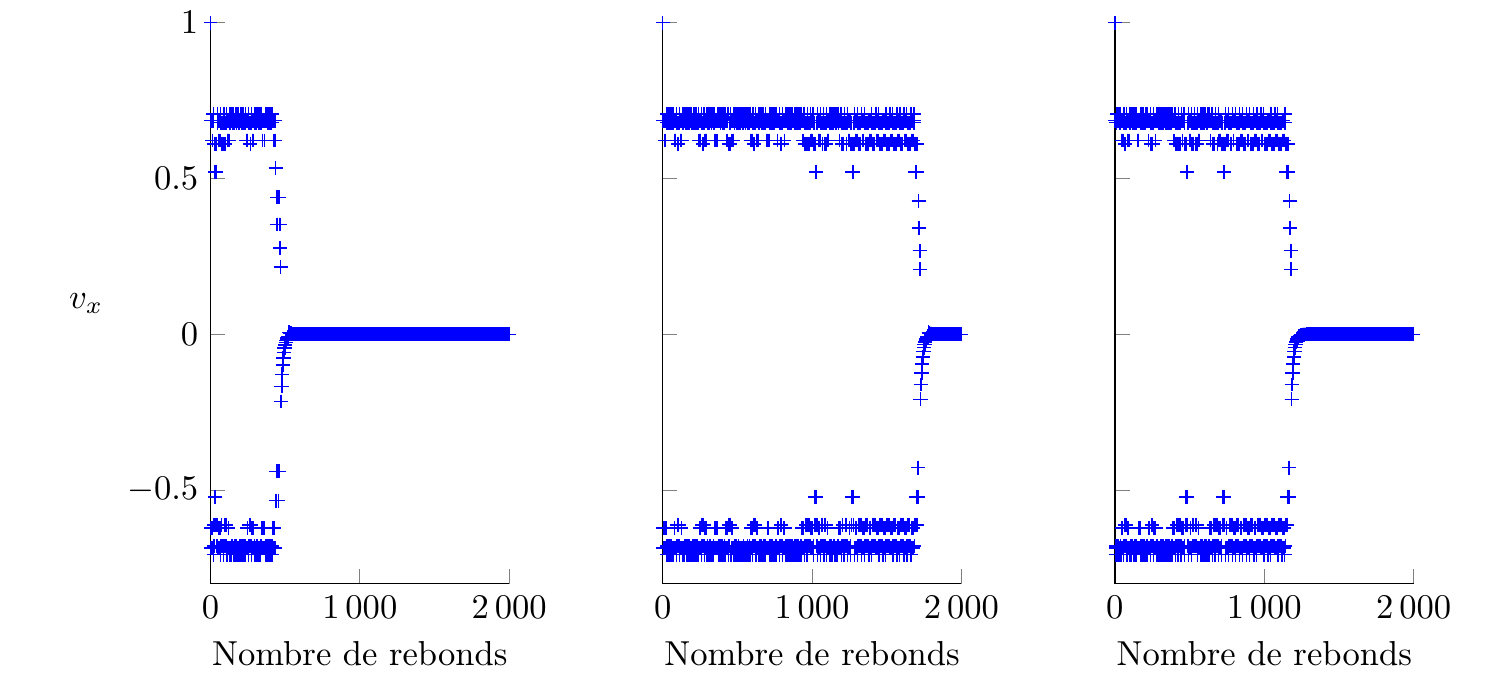}
\caption{Three examples for the evolution of the horizontal component $v_x$ with $y_0=0.1W$ (left panel) $y_0=0.5W$ 
(centre) $y_0=0.85W$ (right). Geometrical parameters are $H=350$, $L=1000$, $W=400$, $\alpha=6\degree$, $\theta=38.3\degree$, $z_0=0.2H$, $x_0= 0$ and $\phi_0= 90\degree$.}
\label{ex_vx}
\end{center}
\end{figure}

These examples show that, due to reflections on the boundaries, the velocity component changes its sign several times,
before the exponential trapping towards zero occurs, as discussed in subsection~\ref{trappingtime}
It is clearly apparent in these examples that the exponential decay lasts much less than the first phase of the evolution.
The trapping time discussed earlier is therefore not always the appropriate quantity to characterize the convergence.
The first regime can last a transitory but long time, before the ray falls into a funnel and becomes fully trapped. 

In summary, despite a value of $s$ being close to 1 that suggests a fast decay of the velocity component $v_x$, the trapping time may be very long, depending 
on the geometrical parameters and initial conditions. Such an effect will be central when dissipation will come into play. Indeed, a large number of 
reflections usually means a long distance of propagation and therefore a strong decay in amplitude when viscous effects can not be neglected before any trapping can take place. In such cases, predictive aspects of ray dynamics become less meaningful.

\subsubsection{Trapping plane}

The longitudinal coordinate  of the trapping $yz$-plane is clearly also an important 
quantity, especially for what concerns any tentative experimental application.  
To determine its coordinate that we will call $x_{\infty}$, one has to study its dependence
with the two launching coordinates $y_0$ and $z_0$. Taking advantage of the
exponential relaxation, the criterion chosen is that the $x$-component
of the velocity field is four orders of magnitude smaller than the two other components.

Figures \ref{pls_attract}(a) and~\ref{pls_attract}(b) show for different initial conditions, 
varying the coordina\-tes~($y_0,z_0$) of the initial launching point (left panel) or of the initial horizontal angle~$\phi_0$ (right panel), the corresponding final paths. All rays converge towards the same structure, an attractor, but whose $x$-coordinate depends
moderately on  $y_0$ 
and $z_0$, but strongly on  $\phi_0$.  By considering the limiting case,  $\phi_0 = 0$, one indeed realizes that 
rays  are restricted to the transverse plane and therefore in that case the trapping plane is $x_{\infty}=x_0$. The collection of attractors in these two panels illustrate the notion of an attracting two-dimensional manifold, existing due to along-slope translational symmetry.

\begin{figure}
\begin{center}
\includegraphics[width=1\textwidth]{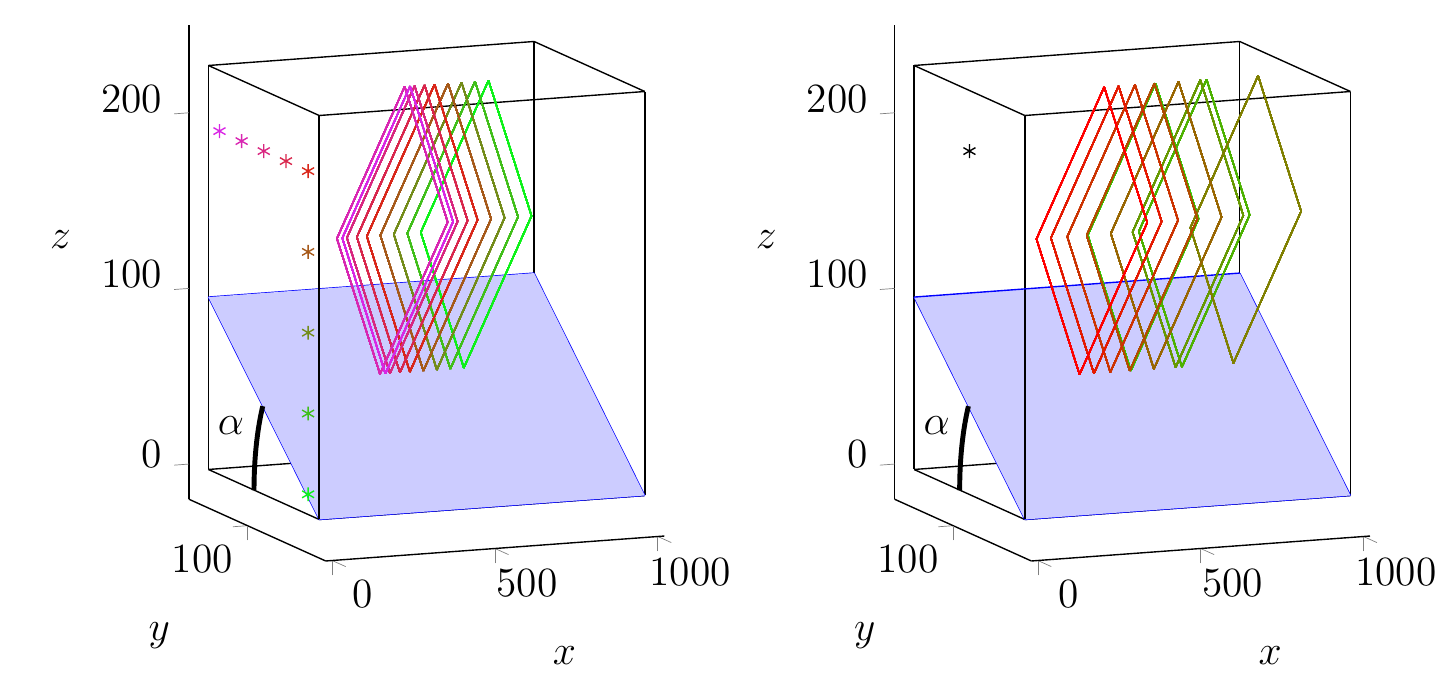}
\caption{The left panel shows final steady paths  for different launching positions in the
$x=0$ plane. The initial launching point and the corresponding attractor are represented with the same colour. $\phi_0$ is always taken equal to $\pi/2$,
The right panel presents results for the same initial launching point (black star) when spanning
values of $\phi_0$ between $20\degree$ ({grey}) and  $90\degree$ ({red}). }
\label{pls_attract}
\end{center}
\end{figure}

Figure \ref{map_foc} presents the position of the trapping plane 
$x_{\infty}$ as a function of $y_0$ and $z_0$ in two cases: a beam propagating initially upward, $v_z = 1$ (left panel), or downward, $v_z=-1$ (right panel). 
The convergence of the first case $v_z = 1$ is in general slower since the first reflection on the sloping bottom
is delayed with respect to the case $v_z = -1$. Other effects may come into play and modify the map. Indeed, discontinuities
 are due to reflection off the end wall of the canal-like geometry  at $x=L$, that delays the trapping. 
These examples show the richness of this dynamical system and emphasize that, even in a case leading to a simple
(1,1) attractor, the convergence towards the trapping plane can be more complicated and with surprises.

\begin{figure}
\begin{center}
\includegraphics[width=0.8\textwidth]{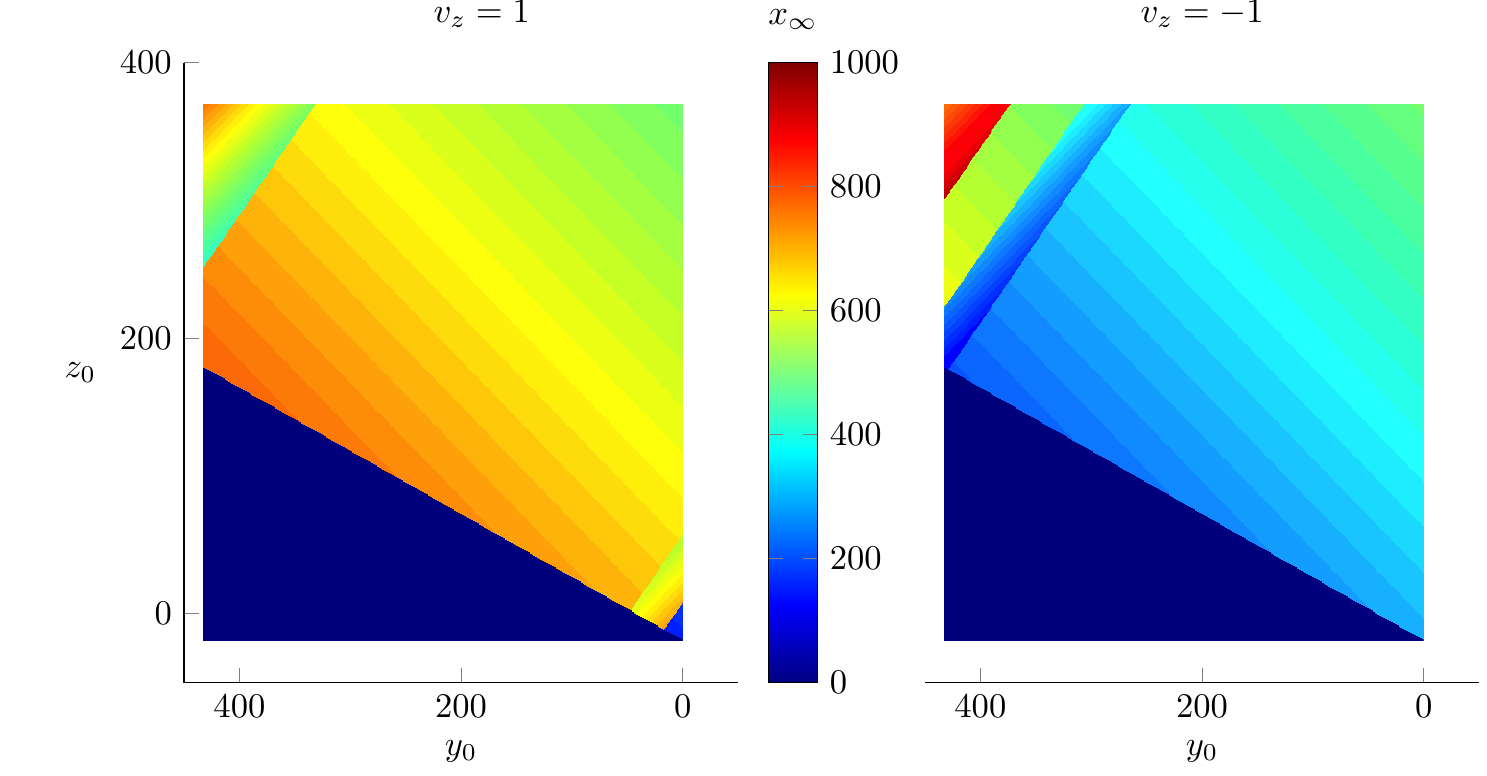}
\caption{Position $x_\infty$ of the trapping plane for different initial conditions   $y_0$ and $z_0$,
while other parameters are kept constant, {particularly $x_0 = 0$ and $\phi_0 = 47 \degree$}.
The left (resp. right) panel corresponds to $v_z=1$ 
(resp. $v_z=-1$). The blue triangle corresponds to the
region below the slope.}
\label{map_foc}
\end{center}
\end{figure}

\subsection{More complicated attractors}
\subsubsection{Phase diagram}

Considering cases not leading to the simple (1,1) attractors,
let us plot for the same geometry the diagram as a function of $\theta$ and $\alpha$. The dimensions $H=360$, $L=1000$ and $W=410$ are constant once the canal is given.
While this  diagram is not universal as the so-called ($d$,$\tau$)-diagram reported for two-dimensional attractors 
in a trapezoidal domain by \cite{Maas2005} and discussed more recently 
in \cite{Arnold2017}, we will show that it allows to vizualize the main regions with simple attractors.
 
Moreover, as discussed above, initial values for the launching point or the horizontal angle are
generically not important since one gets eventually always the same attractor, 
only its position~$x_{\infty}$ changes.
The phase diagram plotted in~\fig{diag_exemple} shows on the left (resp. right) the number of
reflections $m$ (resp. $n$) of the final steady paths on  the vertical wall  $y=0$ (resp. on the slope). They are plotted only for $\alpha < \text{arctan}\, (H/W) \simeq 41 
\degree$, since for larger values the trapezoidal geometry is modified into a triangular domain in which all attractors boil down to a point attractor. 
Both pictures show to what kind of attractors the final steady paths belong to. 

Let us describe the main areas in the diagram:
\begin{itemize}

\item[i)]  The top right {dark} blue triangle corresponds to the convergence towards a domain without reflection,
on the  surface. It corresponds to the point attractors, that one precisely encounters in the supercritical
regime $\alpha > \theta$ (see the central and right panel of~\fig{sketch_attract} in which the ray eventually reaches the bottom right corner of the domain).

\item[ii)]  The blue tongues, present in both the right and left panels correspond to one reflection on the slope and one reflection on the vertical wall. These two pieces of information allows us to conclude the path is the one of a (1,1) attractor.  
Panel (a) of~\fig{diag_exemple} shows  an example of such cases that we discussed in detail in previous sections. 
\item[iii)]  Tongues that are only blue on the left panel (i.e. with  one reflection on the vertical wall, $m=1$) but differently
 coloured (multiple reflections on the slope) on the right panel, 
correspond to $(1,n)$ attractors, with $n$ given by the associated color of the tongue in panel (f).  
Panel~(c) of~\fig{diag_exemple} shows such an example corresponding to a (1,3) attractor. 
\item[iv)] One can identify other structures in the phase diagrams~in Figs.~\ref{diag_exemple}(b) and~(d), and especially,
colored tongues in both panels attesting more complex ($m,n$) attractors.  One (3,1) attractor (panel (b)) and one 
(1,2) global resonance case (panel d),
with 2 reflections from the slope, one focusing and one  compensating defocusing reflection, which occupy a line in panels (e) and (f).
Global resonances have $n$ focusing reflections exactly compensated by  $n$ defocusing reflections. These are characterized by each ray being periodic, instead of, as for attracting cases, approaching a limit cycle.

\item[v)] The remaining large domain in red, and therefore with $(m,n)$ $> 10$, corresponds to even more complicated attractors. 
As for the ($d,\tau$) diagram for two dimensional attractors in a trapezoidal domain, except for singular values (lines), 
one gets attractors for all  ($\alpha$, $\theta$) values.

\end{itemize}

\begin{figure}
\begin{center}
\includegraphics[width=1\textwidth]{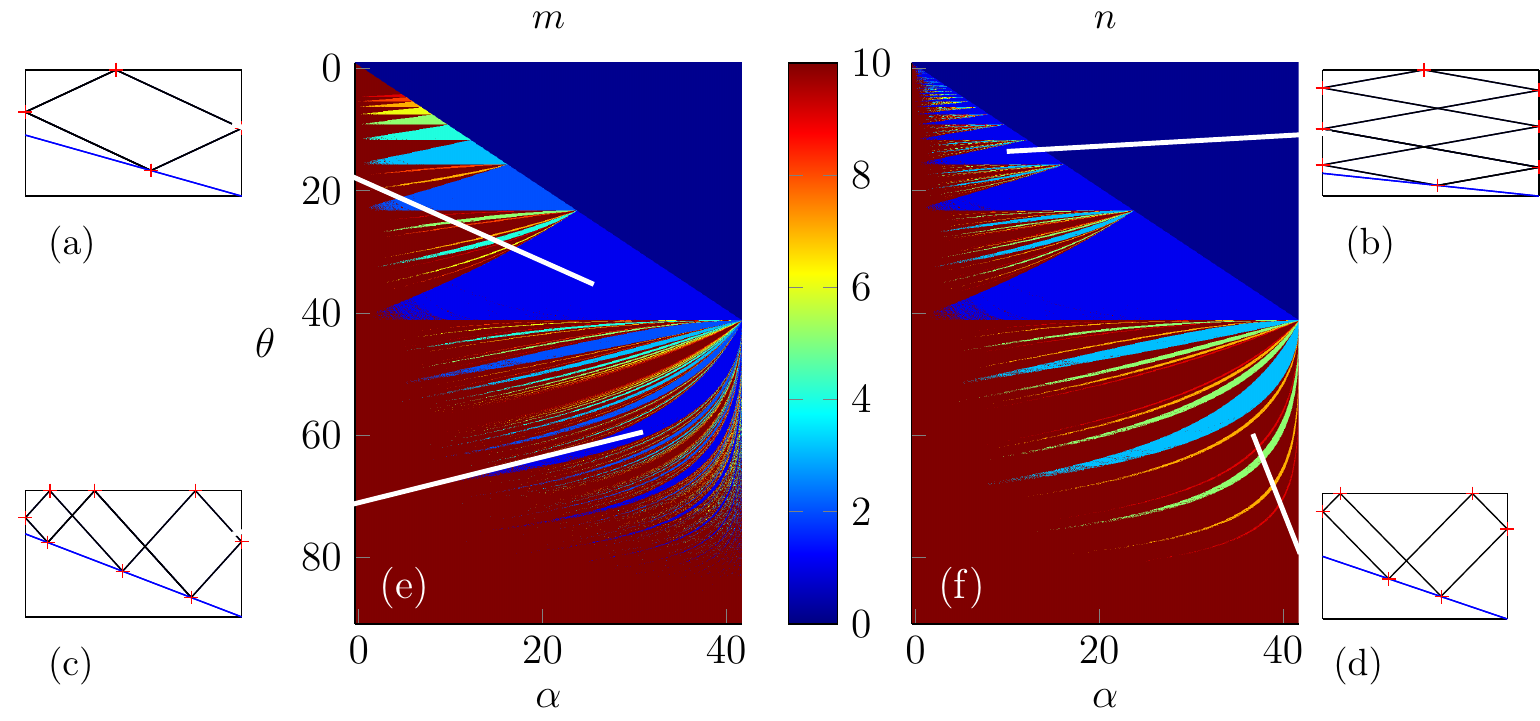}
\caption{Panel (e) and (f): phase diagrams of the steady paths as a function of the angles $\alpha$ and~$\theta$,
for the typical case $H=360$ and~$W=410$, and $L=1000$. 
Panel (e) and (f) give respectively the number of reflections 
off a vertical wall, $m$, and off the slope, $n$, using the central colour table. The top-right triangles correspond to the 
point attractor zone, that exists in the subcritical case.  Three different attractors are shown in 
panel (a), (b) (c) and a global resonance in panel (d), with a link to their region of existence indicated by the white segments.}
\label{diag_exemple}
\end{center}
\end{figure}
It is important to emphasize that there are no attractor regions with an even number of reflections $N_s$ off the slope. This is a property already known for 2D attractors~\citep{Maas2005,PilletPhD}. However, as briefly discussed, there exist lines at which $(m,2n)$ global resonances can be found.

\subsubsection{(1,3) attractor}
Studying a  (1,3) attractor allows to better understand how trapping happens and when it cannot occurs. 
It is indeed known~\citep{Maas2005,PilletPhD}  that  (1, $2n+1$) attractors have $n$ defocusing reflections 
and $n+1$ focusing ones. Indeed, in 2D, more \textit{defocusing} than focusing reflections would mean that rays will on average move away  from one another, which would lead to the absence of attractors.

A typical trajectory converging towards a (1,3) attractor is plotted in~\fig{attract_1_3}. Reflections indexed by 1 and 3 are occurring 
along the gradient of the slope that leads to focusing. On the contrary, reflection 2, being in the opposite direction of the gradient, is defocusing.

In the following, we will consider that the ray still not trapped has a path quasi identical to the one of a two-dimensional (1,3) attractor. Such an approximation
is fully justified by the top view shown in the right panel of~\fig{attract_1_3}. After just a few rebounds, 
one can identify three reflections (identified by the red crosses) on the sloping bottom. Their positions slightly change, but not their focusing or defocusing nature.

\begin{figure}
\begin{center}
\includegraphics[width=1\textwidth]{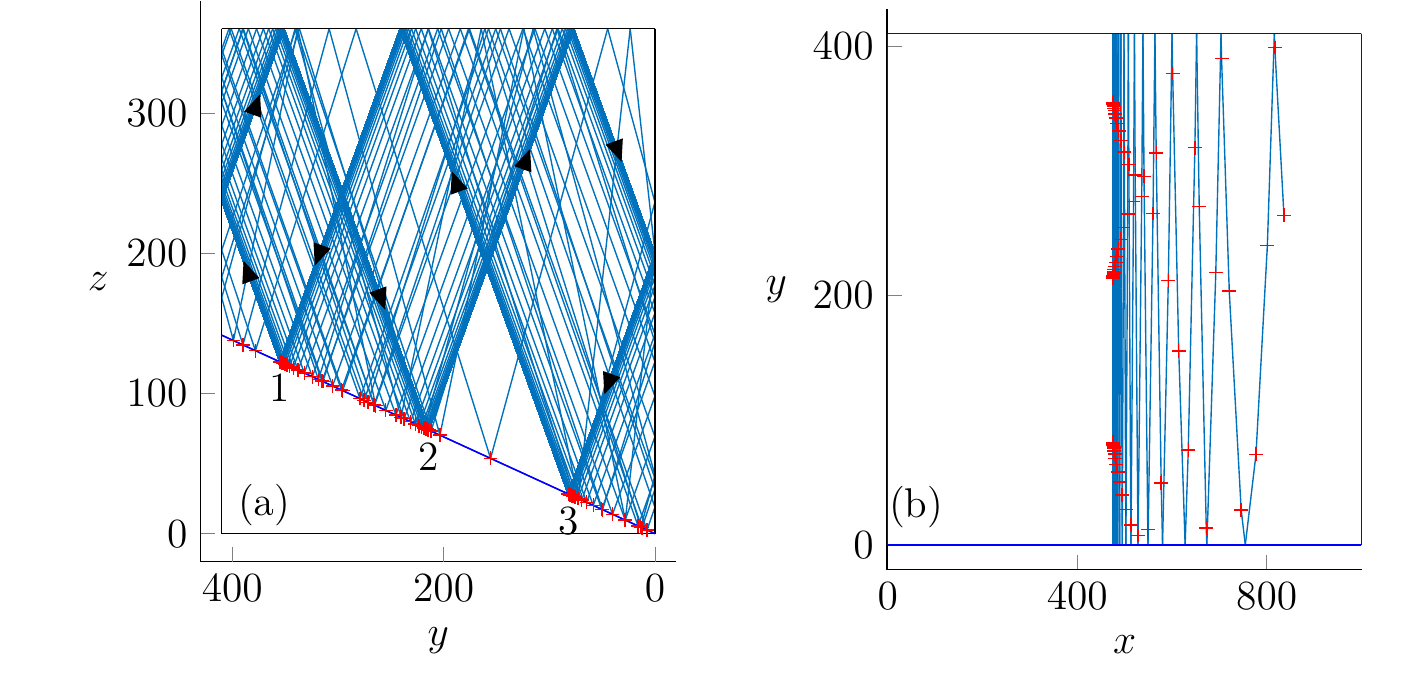}
\caption{Side (a) and top (b) views of a three-dimensional ray tracing in a geometry leading to  (1,3) 	attractor. The corresponding transverse geometry leads to the same (1,3) attractor.  Red crosses locate the reflections on the slope.  Note that for the sake of clarity, the beginning of the ray
tracing,  initiated from $x=0$ {and bouncing back from $x=L$}, has not been plotted. Numbers written just below the slope in (a) indexed the different reflections along the attractor.
}
\label{attract_1_3}
\end{center}
\end{figure}

What can we say for the trapping in such a case? 
One cannot refer to the discussion of~\fig{ptfixe_sur}, in which all reflections led to trapping
since, here, reflections on vertical and horizontal walls do not always give $\phi_i^{'} = \phi_r$.

\begin{figure}
\begin{center}
\includegraphics[width=1\textwidth]{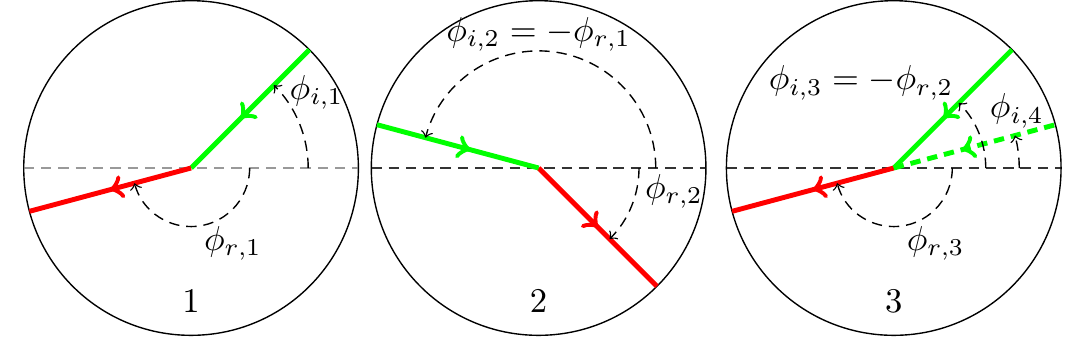}
\caption{Successive reflections on the slope for the trajectory of the (1,3) attractor presented in~\fig{attract_1_3}. Incident
(resp. reflected) rays are plotted in green (resp. red).
}
\label{3horloges}
\end{center}
\end{figure}

Figure~\ref{3horloges} presents the projection of the ray in the $xy$-plane  in the case with three reflections of the (1,3) attractor shown in~\fig{attract_1_3}. Panels (a), (b) and (c) present respectively the reflection numbered 1, 2 and 3. 
Because of the path of the (1,3) attractor, the reflected ray after reflection 1 on the slope will impinge onto the slope
with an incident horizontal angle  $\phi_{i\,2} = 
- \phi_{r\,1}$. The following bottom reflection gives also $\phi_{i\,3} =  - \phi_{r\,2}$. 
One thus realizes that, with respect to reflection 3, reflection 2 leads to an increase of the angle~$\phi$, 
 contributing to untrapping. Consequently,
the normalized velocity component~$v_x$ does not converge anymore monotonically towards~$0$, as shown in~\fig{attract_1_3_vx}.
As Fig.~\ref{3horloges} shows, $\phi_{i,4}=\phi_{i,3}' = \pi+\phi_{r,3} < \phi_{i,1}$ testifying the net focusing after three reflections.
\begin{figure}
\begin{center}
\includegraphics[width=0.7\textwidth]{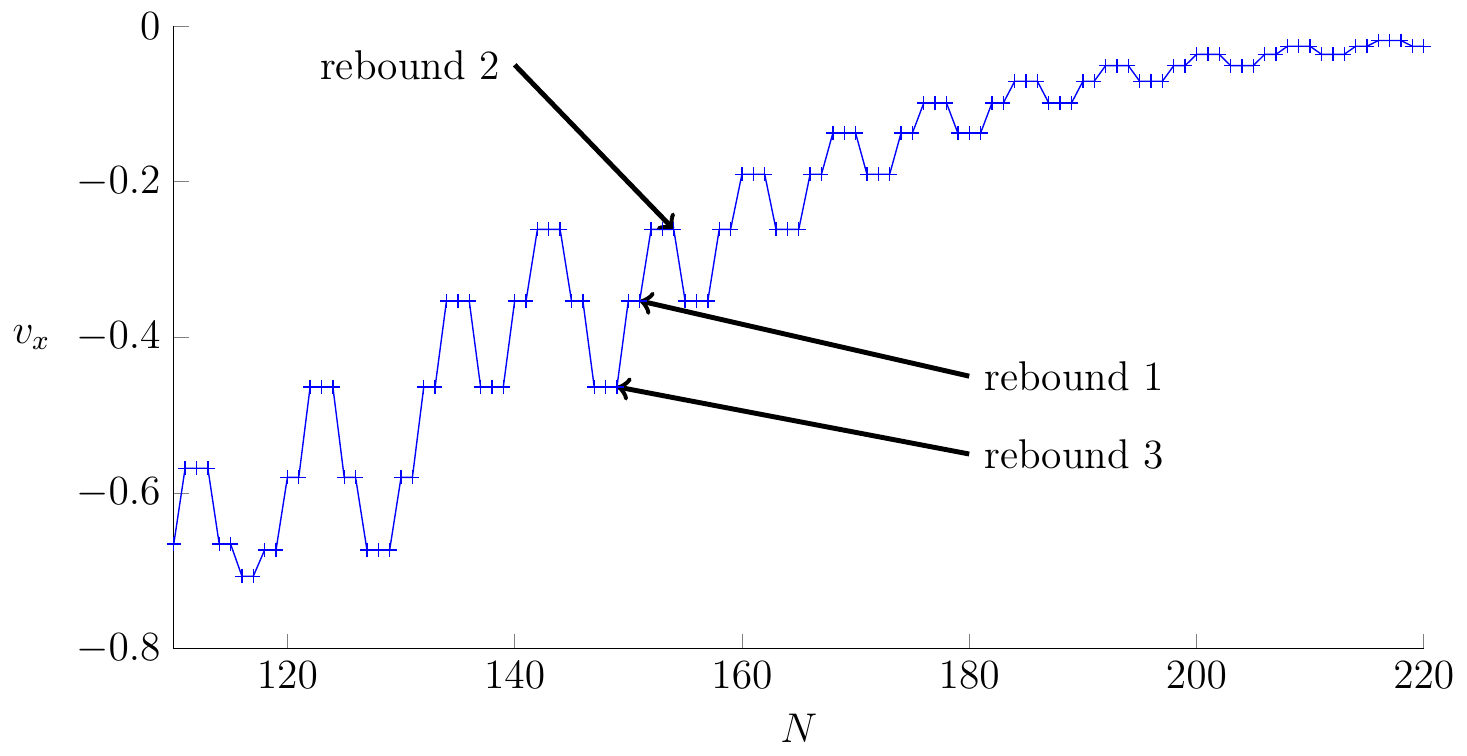}
\caption{Normalized velocity component $v_x$ as a function of $N$ the number of reflections for the trajectory depicted in~\fig{attract_1_3}. 
Two successive rebounds, 1 and 3, lead to a decrease of $v_x$ in absolute value while, on the contrary, rebound 2 leads to an
increase of $v_x$ in absolute value.}
\label{attract_1_3_vx}
\end{center}
\end{figure}

Apart from reflections on vertical or horizontal boundaries, that do not affect~$v_x$, 
one detects two kinds of reflections. Reflections 1 and 3, that lead to a decrease of $v_x$ in absolute value
and reflection 2 that, on the contrary, leads to an increase. Having twice more focusing than defocusing
reflections, the angle converges nevertheless
towards the fixed point $\phi^\star=0$.

The combination of reflections 1 and 2 compensate exactly and leave the velocity component  $v_x$ unchanged, contrary to reflection~3,
the inverse of the convergence time $\gamma_p$ is given by
Eq.~(\ref{gamma}) multiplied by a factor 1/3; only one third of the reflections lead
to a decrease of the velocity.

As in the two-dimensional case, attractors exist generically for any parameter values. Although some singular values 
do not lead to attractors, one cannot get them experimentally or numerically anyway.
As we shall see, one can however not claim that trapping will always occur in 3D since,
as already  discovered for the (1,3) attractor, {trapping and ensuing focusing} can be significantly slowed down and even sometimes not
occur at all. 
The example of (2,1)  global resonance discussed above is an example of such a case.

\subsection{Non trapping cases}
\label{nonpieg}

\subsubsection{$(m,2)$ attractors}

As shown by Fig.~\ref{attract_1_3}, a series of reflections 1-2-...-1-2 would
give rise to a  $(m,2)$ attractor ... that does not exist. Indeed, in two dimensions,
two reflections on the sloping bottom, one focusing while the other one is symmetrically defocusing,
cannot lead towards a limit cycle. On average, rays do not move away from each other:
the Lyapunov exponent is zero. It is known~\citep{Maas2005,PilletPhD} 
that such a case appears only for singular parameter values.
One can however choose to be as close as possible to such a singular point.

\begin{figure}
\begin{center}
\includegraphics[width=1\textwidth]{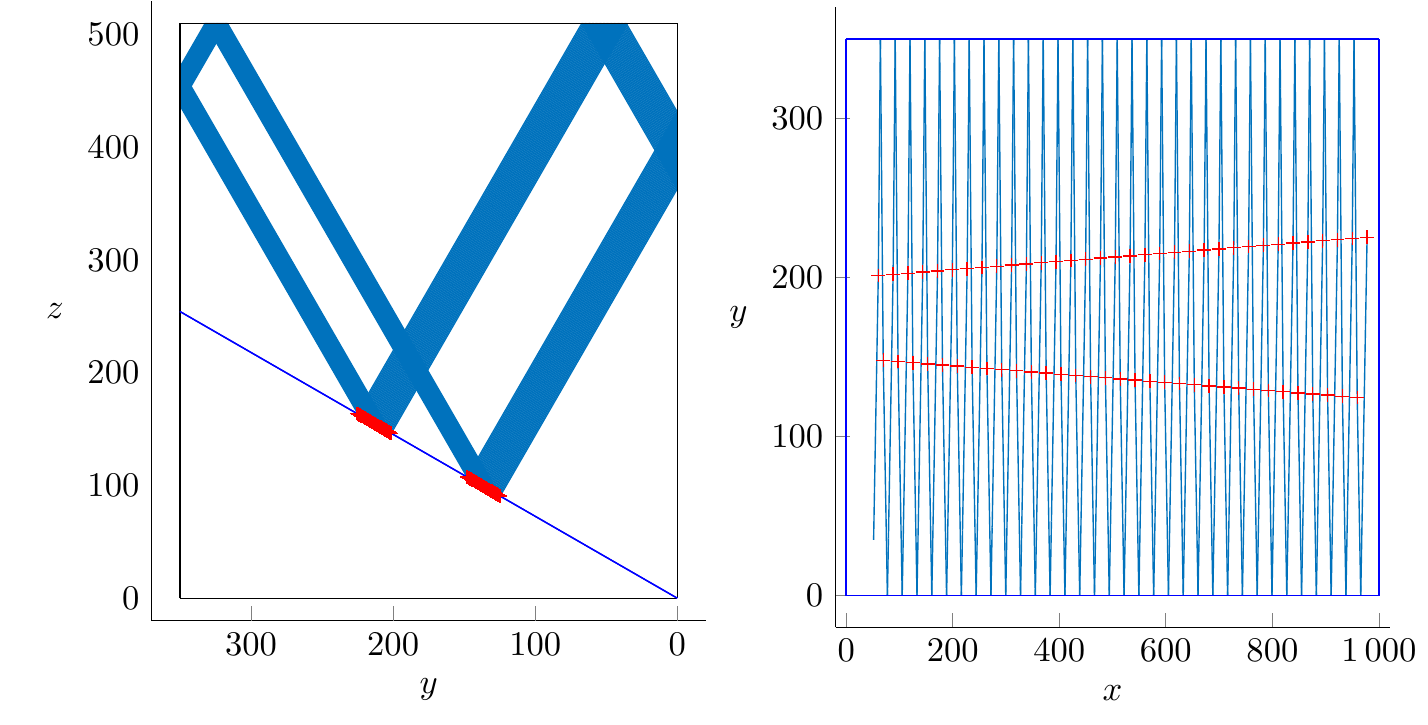}
\caption{Side (left panel) and top (right panel) views of a trajectory for a geometry close 
to a (1,2) attractor that would lead to a vanishing Lyapunov. Reflections on the slope
are identified by red crosses.}
\label{attract_1_2}
\end{center}
\end{figure}

Figure~\ref{attract_1_2} shows that one gets a (1,2) attractor-like structure. 
This is actually not a real attractor: First, trapping does not occur,
which means that there is no limit cycle, and therefore no possible convergence towards it.
Second, this structure is not a steady state. Indeed, after a sufficiently large number
of reflections, the ray converges towards a more complicated true attractor (with more focusing than
defocusing reflections). However, for these values,
the trapping is very slow since the numbers of focusing and defocusing reflections 
are approximately identical. The limit cycle corresponds indeed to  $n$ focusing reflections 
and  $n-1$ defocusing ones, with $n$ large. This is therefore a situation close
to the (1,3) attractor, with a much weaker convergence. 

\subsubsection{Whispering-gallery modes}

Another structure of interest corresponds
to geometries for which, for some well chosen initial conditions rays may escape.
Similar structures have been identified in a trapezoid,  paraboloid, parabolic channel and  spherical geometry for internal gravity or inertial wave rays \citep{MandersMaas2004,Maas2005,DM2007,Rabitti2014}
and are called ~\textit{whispering gallery modes}  in analogy 
with sound waves.
In the system that we study here, such modes exist for very specific parameters and initial conditions:
trapping reflections have to be compensated exactly by  untrapping ones

\begin{figure}
\begin{center}
\includegraphics[width=0.7\textwidth]{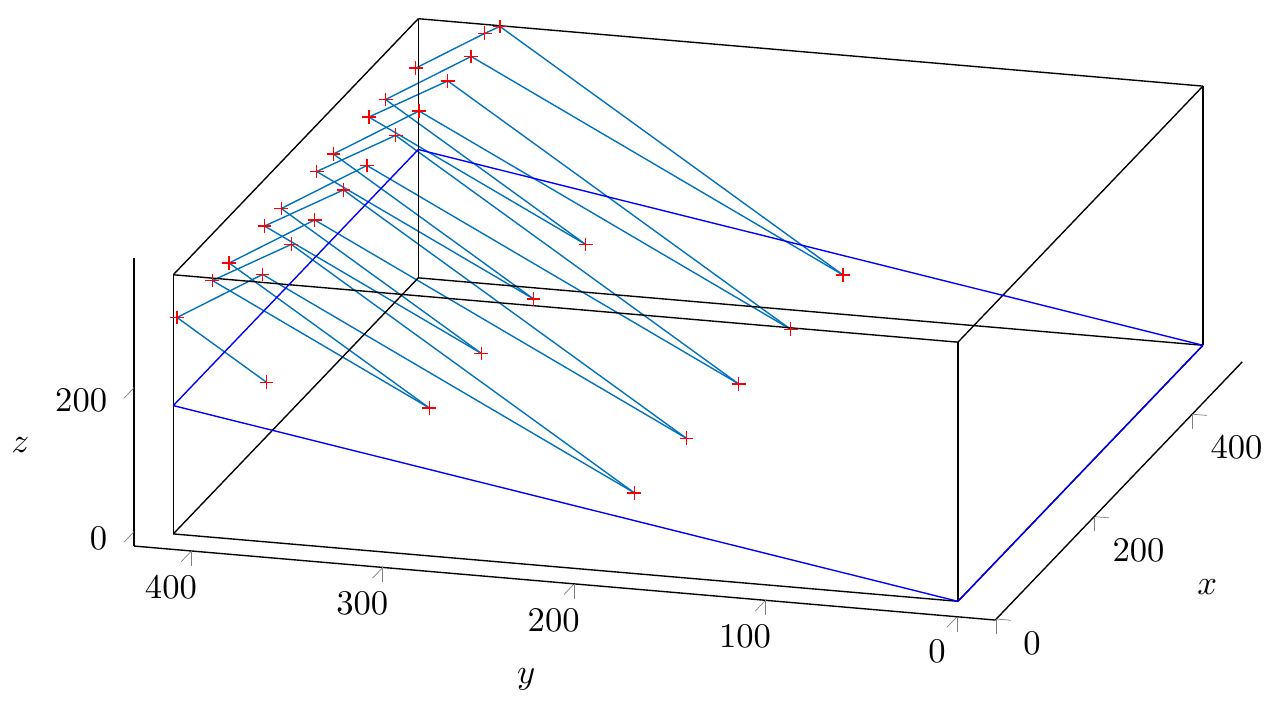}
\caption{Left panel:  trajectory of a  \textit{whispering-gallery 
 mode} with the initial conditions $\phi_0 = 122.476\degree $, $x_0 =0 
$, $y_0=320$, $z_0=324$ and the geometrical parameters
$H=360$, $L=500$, $W=410$, $\theta=39\degree$, $\alpha=23.52\degree$. 
In two-dimensions, this geometry corresponds to the values,
($d=0.1$, $\tau=1.84$) that leads to a (1,1) attractor.
The reflection on the sloping bottom
changes $\phi$ into $- \phi$. After the reflection on the surface and side wall, $y=W$, 
the successive rays will again impinge onto the slope with $\phi_i' = \phi_i$.}
\label{whisper_1}
\end{center}
\end{figure}

Figure~\ref{whisper_1} shows the trajectory of one ray 
in a case that  {(under normal incidence, $\phi_0=0$)}  should lead to a (1,1)  attractor. One can identify a trajectory that is not trapped.
It does not visit the full width of the tank, but stays concentrated on one side of the canal.
A careful look at the values of $\phi$ shows that it stays constant, $\phi=\phi_w$ say, if one forgets symmetries 
with respect to~$x$ and $y$. The value of $\phi_w$ corresponds to the case for which
$\phi_r = 
 -\phi_i$ that leads, as shown in~\fig{whisper_1}, to $\phi_i^{'} = \pi {+}\phi_r = {\pi} -\phi_i$, that explains the stationary state.
Using Eq.~(\ref{loi_sin}), the equality $\phi_r = 
-\phi_i$ corresponds to 
\begin{eqnarray}
\dfrac{s^2-1}{1 + s^2 + 2s\cos \phi_w}  & = &  -1,
\end{eqnarray}
that one can simplify in~$\phi_w = \pi - \arccos s$ 
that precisely correspond to the ray tracing value shown in~\fig{whisper_1}.

Studying the position $x_{\infty}$ of the trapping plane  allows us to identify the existence of whispering-gallery modes that correspond to initial conditions
 that do not converge.  Using the property, shown in~\fig{whisper_1},
 for which the trajectory does not hit the $y=0$ vertical wall, we are able to make the difference
 with 
  trajectories that have not converged yet. 
Figure~\ref{map_whispering} presents the result for different initial conditions 
spanning values of $y_0$ and $\phi_0$ (values of $x_0$ and $z_0$ appear to be much less important).
The two different panels correspond to $x_{\infty}$  after two different numbers of reflections.
Trajectories that have not converged are identified by the white domains, while coloured domains
correspond to different positions of the trapping plane.

\begin{figure}
\begin{center}
\includegraphics[width=1\textwidth]{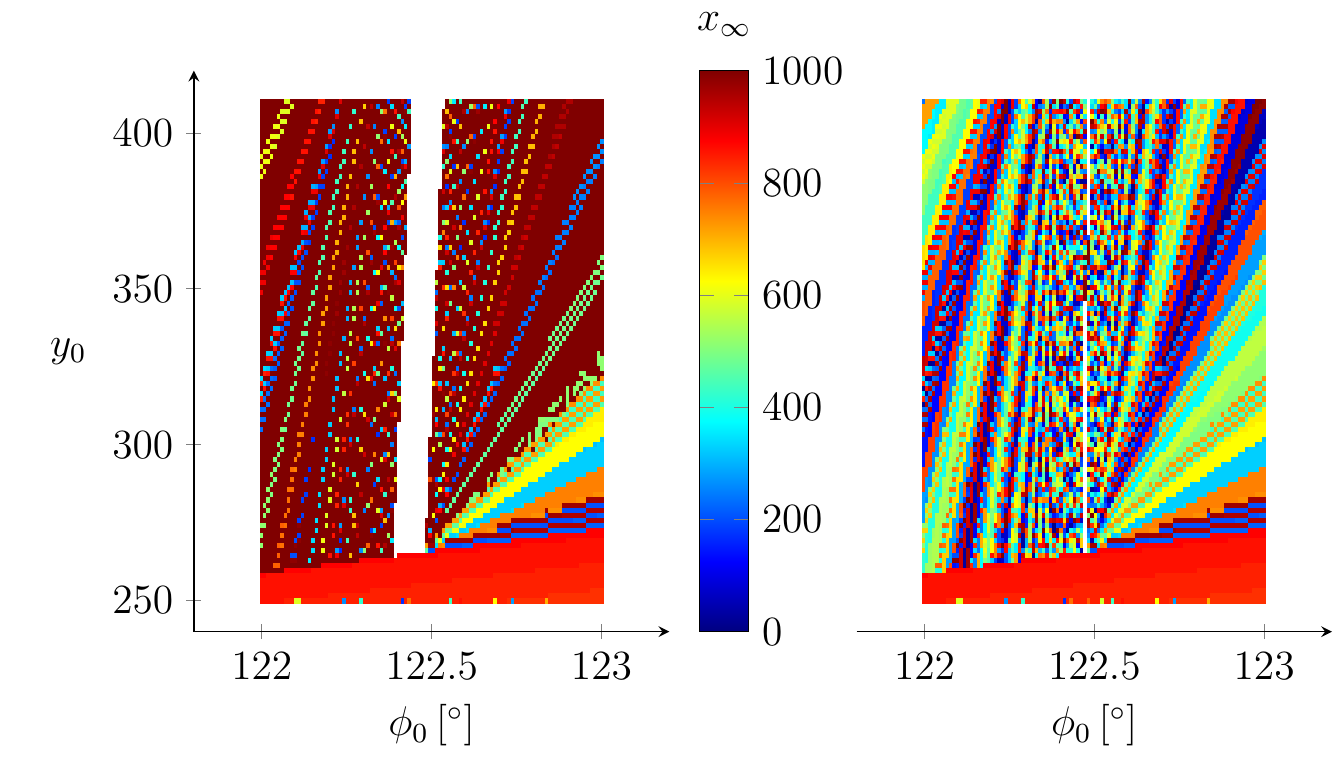}
\caption{Position~$x_\infty$ of the trapping plane for different initial conditions 
 $y_0$ and $\phi_0$. Left (resp. right) panel presents the value
 after $N=10^4$ (resp. $N=10^6$) reflections. White regions that correspond
 to domains in which the iteration has not converged correspond to the~\textit{whispering-gallery modes}.}
\label{map_whispering}
\end{center}
\end{figure}

On the left panel of  \fig{map_whispering}, one sees that  the domain  with whispering-gallery modes (white)
is very thin: $\phi_0 \in[ 122.4\degree,122.5\degree]$.  The right panel shows that this zone has 
drastically shrunk even more if one allows 100 times more reflections: most of this domain is
therefore not  associated with whispering-gallery modes but leads to convergence towards an attractor. It appears finally,
that true  whispering-gallery modes exist only for singular initial conditions. This is after all, 
consistent with the theoretical prediction that only one singular value  $\phi_w$ has been found.

The different results presented in this section have emphasized the links between
the trapping and the existence of attractors in the two-dimensional transverse geometry.
Indeed, as soon as focusing reflections win over defocusing ones in the transverse
two-dimensional plane, three-dimensional reflections lead eventually to trapping.
As in the geometry under scrutiny, if we omit singular values, attractors exist for any angles ($\alpha$, $\theta$) 
and any values $H$ and $W$: trapping will occur generically.
Those singular values are however important. 
Indeed, as attractors are leading to the focusing of energy, dissipation will significantly reduce their energy. In a given geometrical domain,
if the energy is injected in a continuous band of frequencies (i.e. different angles of propagation), the energy will eventually
remain in those that are least dissipated, and therefore in those for which 
focusing does not occur.      

Similarly, although whispering-gallery modes are also the exception rather than the rule, they 
may be visible in a long enough canal in which the energy in the other frequencies
(i.e. for different angles) will be trapped and dissipated in attractors, again because of the focusing mechanism.  
A far enough measurement  in the canal would finally exhibit energy only for
frequencies that were not trapped, and therefore not dissipated before. For this reason they have also been termed 'leaky edge waves'~\citep{DM2007}: despite their low probability, 
if they have been excited upstream, whispering-gallery modes will finally show up.

Moreover, experimental measurements will hardly make a difference between
a not-yet-converged structure or a whispering-gallery mode:
in this sense, \fig{map_whispering}(a) is more appropriate than \fig{map_whispering}(b).
Finally, viscous dissipation was not taken into account in the ray tracing. Although being weak
for internal waves that can travel thousands of kilometers~\citep{RayMitchum},
the distance that would actually represent the convergence towards the limit cyle
of the case discussed in \fig{attract_1_2} is far too long to be observed. 
On the contrary, the transitory quasi-attractor 
is therefore more likely to be observed (see \cite{Pilletal2018}).
In conclusion, from the experimental point of view, only fast trapping cases can be detected.

The key point shown in this section, is that once one knows
that there exists an attractor in a two-dimensional geometry,
its three-dimensional generalization, obtained 
by translation of the 2D geometry along an axis orthogonal to $z$,
will inevitably trap rays during their propagation.

In order to consider more realistic bathymetry, we will now study
a tridimensional geometry that cannot be obtained by the translation
of a 2D geometry.   

\section{A fully tridimensional geometry with super-attractors} 

\subsection{Choice of the geometry}

While all geometries that we studied sofar were translationally  invariant along the length of the canal
(or in other studies rotationally invariant), let us now  study a really tridimensional geometry, in which
 the transverse geometry will vary along the canal. 
Figure \ref{geom_3Dtwist} shows the slope, represented by the blue rectangle, that 
has been obtained from the one shown in \fig{geom} 
after an additional rotation with an angle $\beta$ with respect to the $y$-axis.
In addition to its theoretical interest that we will discuss below, such a study is of course closer to a realistic configuration than previous ones.

\begin{figure}
\begin{center}
\includegraphics[width=0.8\textwidth]{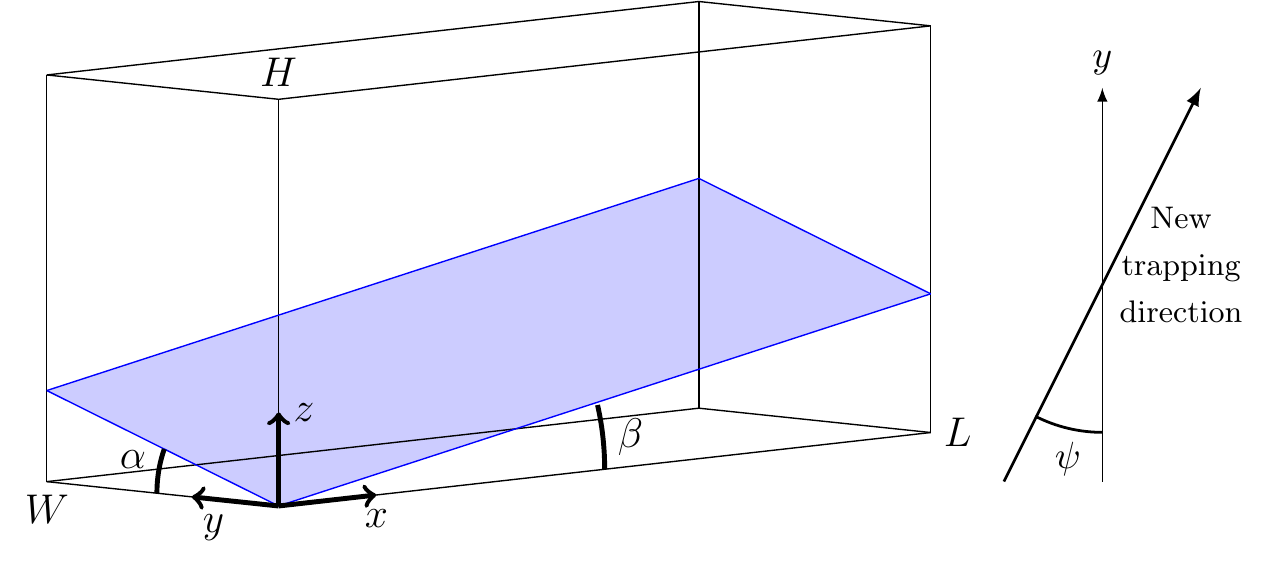}
\caption{Geometry under investigation with the definition of the height $H$, length~$L$ and width $W$ of the canal.
The slope, represented by the blue rectangle, is now inclined with an angle $\alpha$ with respect to the  $y$-axis 
and with an angle $\beta$ with respect to the  $y$-axis.
}
\label{geom_3Dtwist}
\end{center}
\end{figure}

Even if one still uses the angle $\phi$ of a ray with respect to the $y$-axis, one has to
modify the expression of the map  linking the reflected angle $\phi_r$ as
a  function of the incident one~$\phi_i$. Indeed, if the trapping effect has
the tendency to align the angle of 
propagation~$\phi$ along the slope of the gradient, the latter is not along the  $y$-direction, 
but {along} a direction rotated by an angle $\psi$ with respect to the $y$-axis, defined as 
$\tan \psi = -\tan \beta / \tan \alpha$ (see Fig.~\ref{geom_3Dtwist}). 

With this modification taken into account, formula~(\ref{loi_sin}) has to be rewritten as 
\begin{equation}
\sin (\phi_r - \psi)  
=\dfrac{(s^2-1)\sin (\phi_i-\psi)}{(1+s^2) + 2s\cos(\phi_i-\psi)}
\label{calRpsi}.
\end{equation}
The associated map, that we will call ${\cal R}_\psi$, always has a fixed point corresponding to $\phi^* = \psi$.

\subsection{Trapping conditions}

{Despite the new reflection law~(\ref{calRpsi}), trapping will still occur. A wave in  the transverse $yz$-plane
will have the tendency to align with the upslope-directed gradient. 
If, in previous geometries, vertical walls $y=0$ and $y=L$ were oriented perpendicularly to the trapping direction, this is not the case  any more  with this geometry. 
 On the contrary, a wave corresponding to 
$\phi_i = \psi$, will lead {(after three succesive reflections on a vertical wall, the free surface and a vertical wall like in figure \ref{sketch_attract}a) {to a bounce on the slope with $\phi_i'= -\psi$, that will untrap the wave.} 

From this simple remark, one can immediately deduce that two-dimensional attractors are therefore not possible
in this geometry. More generally, it is also straightforward to realize that, in three dimensions, an internal wave can converge
towards a two-dimensional plane only if the upslope-directed gradient 
 belongs to this plane,
while the vertical walls are perpendicular to it.
In the geometry under scrutiny, if the walls are perpendicular to the upslope-directed gradient,
 one recovers the canal geometry studied in the previous section.
One thus gets that two-dimensional attractors can be found only when $\psi=0$, i.e. for $\beta=0$ as studied in Sec.~\ref{canal}.

However, the above remark does not prohibit the existence of attractors in domains having tri-dimensional geometries.
Such issue is far more complicated and we will 
now give some new insight along this line. 

\subsection{Tri-dimensional super-attractors}

Despite the impossibility to get two-dimensional attractors, one can
consider cases for which the transverse geometry is close to the one with a (1,1) attractor in 2D. 
Intuitively, if three dimensional structures do exist, they should {have} a few 
rebounds and be therefore easier to handle. Let us exhibit such a tridimensional structure for a given set
of parameters.  
\begin{figure}
\begin{center}
\includegraphics[width=\textwidth]{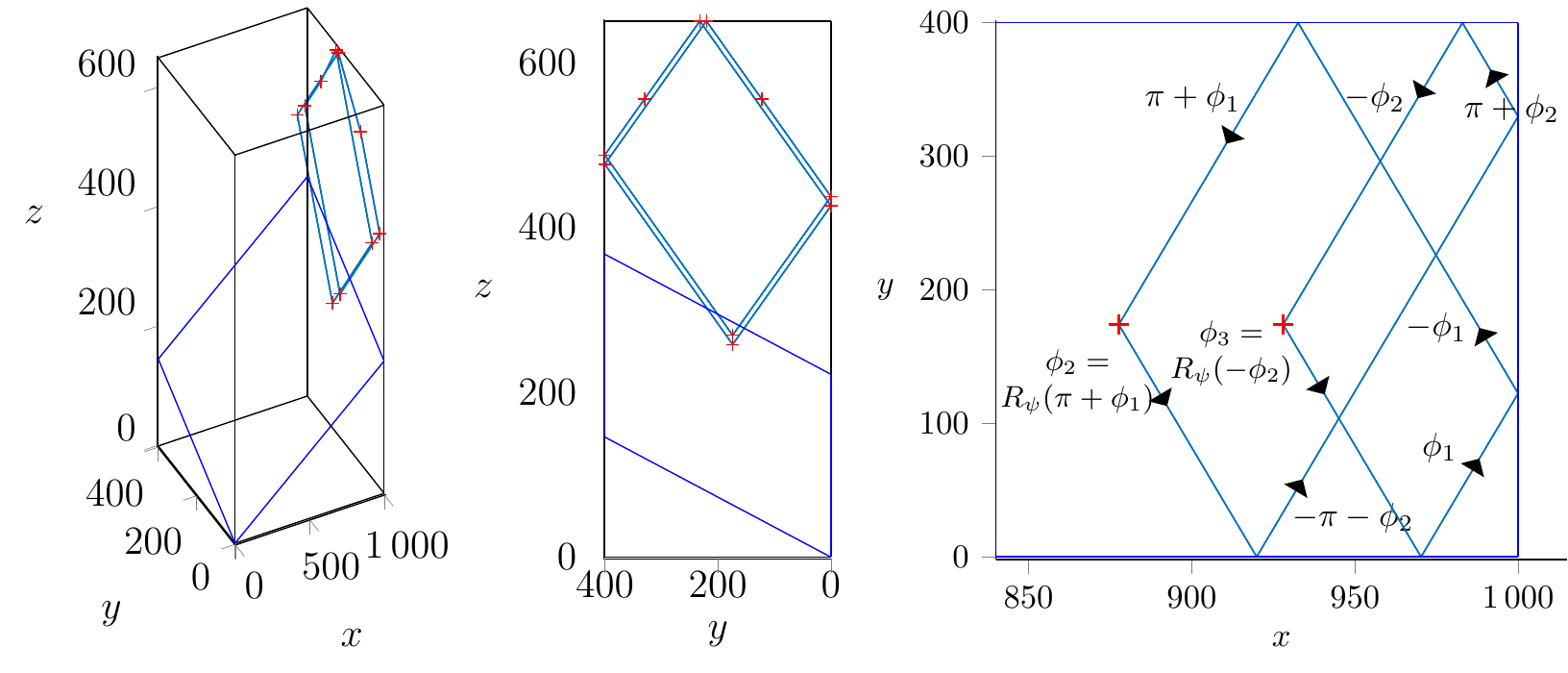}
\caption{Stationary structure in the tri-dimensional case with $H=650$, $L=1000$, $W=400$, $\theta=43.2\degree$, $\alpha = 20\degree$ and 
$\beta=12.5\degree$. The left (resp. centered, right) panel presents a  perspective (resp. side, top) view. 
In the right panel, reflections on the slope are identified by red crosses.}
\label{super3}
\end{center}
\end{figure}

Figure~\ref{super3} presents the stationary structure that one gets with different views. 
As shown by the left panel, the structure is confined close to the end of the canal~$x=L$,
while the side view is strongly reminiscent of a (1,1) attractor. On the contrary, the top view 
emphasizes its three-dimensional nature, with two different reflections on the inclined slope. 

The successive reflections and the different angles of propagation have been plotted on the right panel of 
\fig{super3}. Calling $\phi_1$ the initial angle, the incident beam hits the walls  $x=L$ and $y=W$, 
before impinging onto the inclined slope. The reflected angle is therefore
 $\phi_2 = {\cal R}_{\psi}(\pi + \phi_1)$. After three reflections on the vertical walls  the second reflection
 on the slope occurs, leading to an angle of propagation $\phi_3 = {\cal R}_{\psi}(-\phi_2)$ that has to be equal 
 to $-\pi - \phi_1$ so that the cycle will start again. One thus ends with an angle~$\phi_1$ 
that fulfills the following equation
\begin{eqnarray}
\phi_1 & = & -\pi - {\cal R}_\psi(-\phi_2) \\
 & = & -\pi - {\cal R}_\psi(-{\cal R}_\psi(\pi+\phi_1)).
\end{eqnarray}  
Finding a fixed point of this equation is not easy to exhibit analytically since 
${\cal R}_\psi(-\phi)$ can not be simplified since 
$\psi$ breaks the symmetry with respect to the $y$-axis. 
Using the ray tracing approach, one can find 
the fixed point  $\phi^* \simeq -14 \degree $. 
The next reflection on the vertical wall leads to an untrapping that is exactly compensated 
by the following reflection on the slope. These steps are detailed in Fig.~\ref{super3}c.

Figure~\ref{super_diff} presents the different stationary structures that one obtains by varying either
 $\phi_0$ (left panel) or  $y_0$ and $z_0$ (right panel).
Although leading to different stationary structures, one always gets a limit cycle with two reflections
off the wall at $x=L$, similar to the structure shown in \fig{super3}. Such a behavior is different 
from the attractors obtained in the simpler canal geometry discussed in previous section.  

\begin{figure}
\begin{center}
\includegraphics[width=1\textwidth]{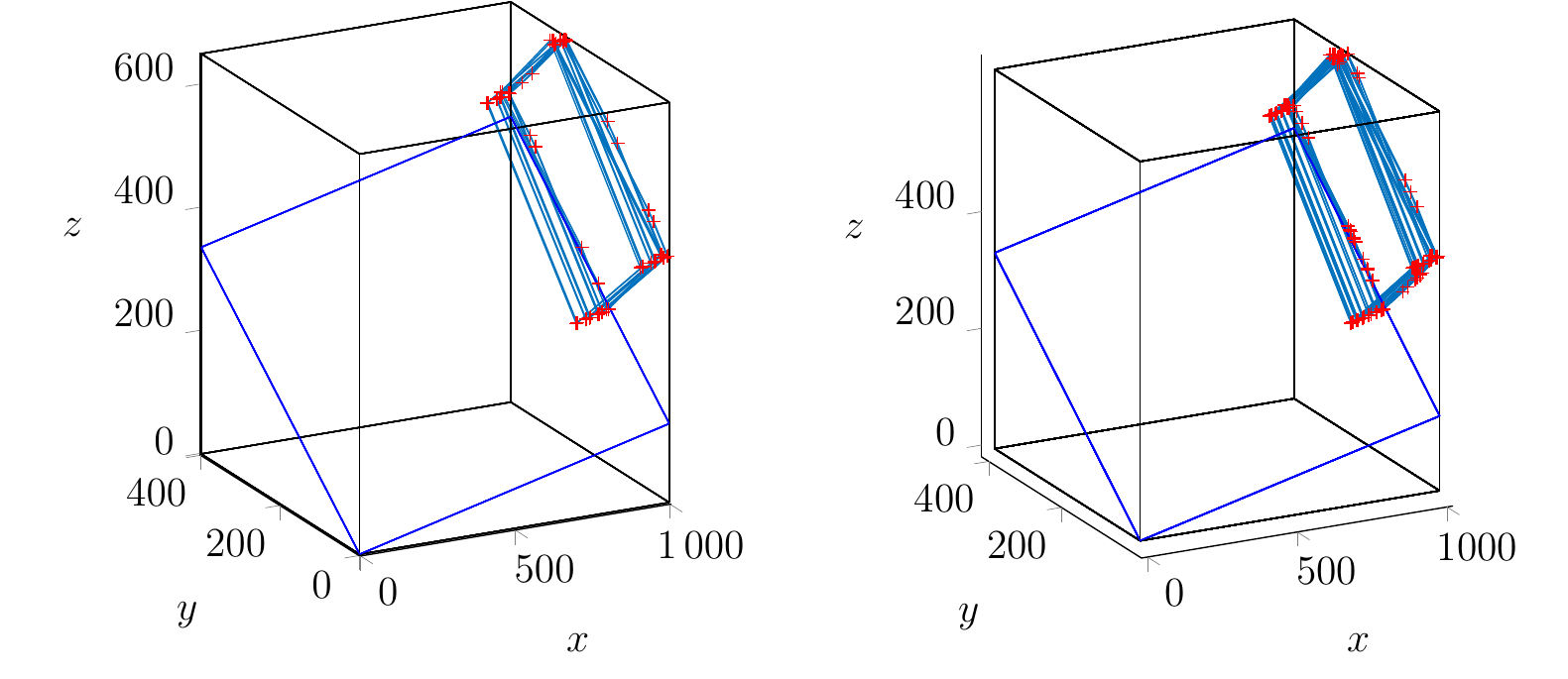}
\caption{Three-dimensional attractors obtained for different initial conditions, varying $\phi_0\in[-80\degree,80\degree]$ (left panel) or varying 
 $y_0\in [ 0,W]$ and $z_0\in[0,H]$ (right panel). All other parameters are those given in the caption of \fig{super3}.}
\label{super_diff}
\end{center}
\end{figure}

Despite the small {variations in the bouncing coordinates}, the limit cycle is almost unchanged for all initial conditions.
One thus gets a situation close to the two-dimensional one, with the difference that they occur
only close to the end wall $x=L$. The change of the geometry in the longitudinal direction of the canal has important consequences:
trapping regions that were possible all along the translationally invariant canal, are now confined to the region
close to the end of a canal that is not translationally invariant. This is therefore a "super-attractor" since the energy is trapped
in {a} significantly more restricted region.

In the example discussed above, only one reflection is {sufficient to make a significant change in the angle of propagation}.
However, for lower values of  $s$ and larger values of~$\psi$, the wave can
be trapped after several reflections. Using similar ideas, it is therefore possible
to exhibit even more complicated structures, with progressive
trapping with reflections on the slope and untrapping on the wall $x=L$~\citep{PilletPhD} .

\section{Conclusion}

In this paper, we have shown that although the dispersion relation is unchanged, the propagation in three dimensions is significantly more complicated than its two-dimensional version. 
The wave propagation on a cone that generalizes the Saint Andrew's cross justifies the introduction of an additional
angle of propagation $\phi$ that allows to describe the position of a wave ray in the horizontal plane.

We have studied the evolution of this reflection angle over inclined slopes and shown the emergence of
a new mechanism that has the tendency to align this angle  $\phi$ with the upslope gradient. 
We have also carefully studied this trapping in the rather simple geometry of a translationally invariant canal.
This configuration leads to a trapezium very similar to what has been extensively studied in two-dimensions.
It is however important to emphasize that we also established a direct link between the trapping and the existence of two-dimensional attractors.
The important feature is that, in such a case, there is not only one attractor that would attract all rays, but an infinity of two-dimensional attractors distributed 
along the canal, that we refer to as a two-dimensional attracting manifold.

We have also considered a geometry that is not translationally invariant which is closer to realistic configurations.
In this new geometry, we were able to prove that there are no two-dimensional attractors. However, we have exhibited
a three-dimensional structure with properties similar to internal wave attractors. Moreover, as it is unique, it is likely that
it should be easy to visualize it in laboratory experiments since the energy injected in the domain would be eventually
confined to a very thin region in three-dimensional space: a \textit{one}-dimensional manifold, which is the reason for calling it a super-attractor.  
The experimental verification of this prediction is one of our priorities. 

As a side remark, we note that  translational invariance is of course broken at the front and end walls of the canal where frictional effects might modify the attracting structures~\citep{Beckebanze2018}. For internal gravity waves, having rectilinear particle motions, this is not seen as leading to major changes. But for the analogous case of inertial waves, that possess (inclined) \textit{circular} particle motions this is an issue, and adjustment at an inviscid level is to be expected. A preliminary experimental study of the attracting two-dimensional manifold of inertial waves does indeed show adjustment of the cross-sectional attractor shape on approach of the side walls ~\citep{MandersMaas2004}.

The study of attractors in three dimensions is however still in its infancy and we expect other very interesting features to discover. 
Considering more complicated attractors in even more realistic configurations is important.
Moreover, the existence and the likelihood of super-attractors in generic three dimensional geometry is fully open
and could lead to interesting predictions when considering the real bathymetry of oceans: that should lead the way 
for observations in the oceans.

\begin{acknowledgments}
{\bf Acknowledgments}
This work was supported by the LABEX iMUST (ANR-10-LABX-0064) of Universit\'e de Lyon, within 
the program ``Investissements d'Avenir'' (ANR-11-IDEX-0007), operated by the French National 
Research Agency (ANR). This work has been supported by the ANR through grant ANR-17-CE30-0003 (DisET).
This work has been achieved thanks to the resources of PSMN from ENS de Lyon. \end{acknowledgments}

\appendix
\section{}
The incident and reflected velocity fields share a reciprocal relationship as they should switch role upon time-reversal. This is borne out when rewriting Eqs. (\ref{neweqa}) and (\ref{neweqb})  as 
\begin{eqnarray}
\begin{pmatrix} 
-s & 1 \\
1 & -s 
\end{pmatrix} 
\begin{pmatrix} 
v_y  \\
v_z  
\end{pmatrix}_r
=
\begin{pmatrix} 
s & -1 \\
1 & -s 
\end{pmatrix} 
\begin{pmatrix} 
v_y  \\
v_z  
\end{pmatrix}_i
\label{mat1}\end{eqnarray}
or, {employing the projection of the velocity vector onto the plane perpendicular to the sloping topography, $\vec{v}_{\perp}=(v_y,v_z)$,}
\[
P\vec{v}_{\perp,r}  = Q\vec{v}_{\perp,i}, \]
where we use unimodular matrices that have determinant $\pm 1$:
\[ P \equiv {1 \over \sqrt{1-s^2} }\begin{pmatrix} 
-s & 1 \\
1 & -s 
\end{pmatrix}, \;\;\quad Q \equiv {1 \over \sqrt{1-s^2}}\begin{pmatrix} 
s & -1 \\
1 & -s 
\end{pmatrix}. \]
Interestingly, these hyperbolic matrices reflect their reciprocal nature by obeying the identity \[P^{-1}Q=Q^{-1}P\equiv R,\]
which implies 
\[\vec{v}_{\perp,r} =R\vec{v}_{\perp,i},\;\;\mbox{and}\;\; \vec{v}_{\perp,i} =R\vec{v}_{\perp,r}.\]
The matrices acquire standard form when, for subcritical slope, $s<1$, we write $s=\tanh \nu$, so that
 \begin{equation} 
 P \equiv 
 \begin{pmatrix} 
-\sinh \nu & \cosh \nu \\
\cosh \nu & -\sinh \nu 
\end{pmatrix}
\quad \mbox{and}\quad
 Q \equiv \begin{pmatrix} 
\sinh \nu & -\cosh \nu \\
\cosh \nu & -\sinh \nu 
\end{pmatrix}.
 \end{equation} 
Notice that $\det(P)=-1$ and $\det(Q)=1$. 

For supercritical topography, $s>1$, we premultiply left and right hand sides of (\ref{mat1}) by  $-1/\sqrt{s^2-1}$, and, writing $s=\coth \mu$, we find
 \[ P \equiv \begin{pmatrix} 
 \cosh \mu & -\sinh \mu  \\
  -\sinh \mu & \cosh \mu
\end{pmatrix}
\quad \mbox{and}\quad
Q \equiv \begin{pmatrix} 
 -\cosh \mu & \sinh \mu  \\
  -\sinh \mu & \cosh \mu \end{pmatrix}.
\]
Defining $Q_n=Q(n \mu)$, such that  the previously defined $Q \equiv Q_1$, it appears that 
\[ P^{-1} Q= Q^{-1}P=Q_2.\]

These reciprocal relations are useful when computing the velocity vector upon a ray's reflection from a boundary, given the incident velocity vector. It also helps determining the proper root when  solving the multivalued (\ref{loi_sin}) for horizontal direction $\phi_r$, and subsequent angle of incidence $\phi_i'=\pi+\phi_r$, given the incident angle of incidence $\phi_i$ and slope $s$.

For subcritically sloping topography, $s<1$, with  
\[\Phi(\phi,s)\equiv \sin^{-1}\left({(1-s^2) \sin \phi \over 1+s^2+2s \cos \phi}\right),\] and 
\[\varphi(s) \equiv  \pi - \sin^{-1} \left( {1-s^2 \over 1+s^2} \right)=\pi-\Phi\left({\pi \over 2},s\right), \] we obtain as subsequent angle of incidence
\[
\phi'= {\cal R}(\phi,s) \equiv \begin{cases}
      \pi-\Phi(\phi,s), & \text{if}\ \phi < -\varphi(s) \\
      \Phi(\phi,s), & \text{if}\ |\phi| < \varphi(s) \\
      \pi+\Phi(\phi,s), & \text{if}\ \phi > \varphi(s)
    \end{cases} , \]
   displayed in Fig.\;\ref{ptfixe_sur}. The conditions apply to secure continuity when $\phi'$ passes $\pm \pi/2$.
   
For supercritical topography, $s>1$, we need to distinguish between rays incident from above and below. 
    Recalling that $\phi_\ell(s) = \tan^{-1} \sqrt{s^2-1}$, for rays incident from above ($v_{z,i}<0$), we find
    \[\phi'= {\cal R}(\phi,s) \equiv -\Phi(\phi,s)\;\text{if}\; -\pi+\phi_\ell(s) \le \phi \le \pi-\phi_\ell(s),\]
    displayed in Fig.\;\ref{ptfixe_sous}a,
    while for rays incident from below ($v_{z,i} > 0)$, i.e. for $-\phi_\ell(s) \le \phi \le \phi_\ell(s)$, we have the reciprocal relation
  \[
     \phi'= {\cal R}(\phi,s) \equiv \begin{cases}
      -\pi+\Phi(\phi,-s), & \text{if}\ \phi <\pi -\varphi(s) \\
      -\Phi(\phi,-s), & \text{if}\ |\phi| < \varphi(s)-\pi  \\
      \pi+\Phi(\phi,-s), & \text{if}\ \phi > \varphi(s)-\pi
    \end{cases} , \]
displayed in Fig.\;\ref{ptfixe_sous}b.

\end{document}